\documentclass[11pt]{article}

\usepackage[utf8]{inputenc}

\usepackage{float}
\usepackage{fullpage}
\usepackage{graphicx}
\usepackage{array}
\usepackage{paralist}
\usepackage{verbatim}
\usepackage{epstopdf}
\usepackage[font=footnotesize]{subfig}
\usepackage[cmex10]{amsmath}
\usepackage{amssymb}
\usepackage{amsthm}
\usepackage{amsfonts}
\usepackage{algpseudocode}
\usepackage{algorithmicx}
\usepackage{fixltx2e}
\usepackage{mdwmath}
\usepackage{mdwtab}
\usepackage{hyperref}
\usepackage{cite}

\theoremstyle{definition}

\theoremstyle{plain}
\newtheorem{theorem}{Theorem}
\newtheorem{lemma}{Lemma}

\newcommand\blfootnote[1]{%
  \begingroup
  \renewcommand\thefootnote{}\footnotetext{#1}%
  \addtocounter{footnote}{-1}%
  \endgroup
}

\title{A meet-in-the-middle algorithm for fast synthesis of depth-optimal quantum circuits}
\author{\small{Matthew Amy$^{1}$, Dmitri Maslov$^{2, 3}$, Michele Mosca$^{4,5}$, and Martin Roetteler$^{6}$} \\
{\small\it $^1$ Institute for Quantum Computing, and David R. Cheriton School of Computer Science} \\
{\small\it University of Waterloo, Waterloo, Ontario, Canada} \\
{\small\it $^2$ National Science Foundation} \\
{\small\it Arlington, Virginia, USA} \\
{\small\it $^3$  Institute for Quantum Computing, and Dept. of Physics \& Astronomy} \\
{\small\it University of Waterloo,  Waterloo, Ontario, Canada} \\
{\small\it $^4$  Institute for Quantum Computing, and Dept. of Combinatorics \& Optimization} \\
{\small\it University of Waterloo,  Waterloo, Ontario, Canada} \\
{\small\it $^5$ Perimeter Insitute for Theoretical Physics} \\
{\small\it Waterloo, Ontario, Canada} \\
{\small\it $^6$ NEC Laboratories America} \\
{\small\it Princeton, New Jersey, USA} \\
}

\begin{document}

\maketitle

\blfootnote{This work has recently been accepted by IEEE for publication. Copyright may be transferred without notice, after which this version may no longer be accessible.}

\addtocounter{footnote}{1}%

\begin{abstract}
We present an algorithm for computing depth-optimal decompositions of logical
operations, leveraging a {\em meet-in-the-middle} technique to provide
a significant speed-up over simple brute force algorithms. As an
illustration of our method we implemented 
this algorithm and found factorizations of the commonly used quantum logical operations into 
elementary gates in the Clifford$+T$ set. In particular, we report a decomposition of the
Toffoli gate over the set of Clifford and $T$ gates.  Our decomposition achieves a total $T$-
depth of 3, thereby providing a 40\% reduction over the previously
best known decomposition for the Toffoli gate. Due to the size of the search space the
algorithm is only practical for small parameters, such as the number of qubits, and the number of gates in an optimal implementation.
\end{abstract}

\section{Introduction}

In classical computing, CPUs typically have a small, fixed set of instructions for operating directly on the words in memory. However, it is much more convenient when writing complex functions to use higher level operations which correspond to (potentially long) sequences of CPU instructions. A programming language provides such operations, and it is then a compiler's job (amongst other things) to expand the higher level code into CPU instructions. While the compiler will most likely do some optimizations both before and after this process, there are some standard sequences of instructions that operations in the programming language get expanded to.

A quantum computer faces similar difficulties -- though in fact, the necessity for a fixed set of instructions is even more pronounced, due to fault tolerance protocols and error correction. The fault tolerance protocols also greatly affect how efficient the given instructions are, mirroring the structure of various assembly languages, notably ARM, where one instruction may correspond to widely varying numbers of clock cycles. At the higher level however, most quantum algorithms are described using a wide variety of gates, and given a specific instruction set consisting of a few fault tolerant logical gates, a compiler would need to know how to implement these gates efficiently using the instruction set. Given that many logical operations are commonly used in quantum algorithms, these gates can and should be factored optimally off-line.

In order to best exploit the limited quantum computational resources available, it will be most important to reduce the resources required to implement a given quantum circuit. With recent advances in quantum information processing technologies (e.g. \cite{QS}, \cite{SQ}, \cite{IBM1}, \cite{IBM2}) and improvements to fault-tolerant thresholds (e.g. \cite{B1}, \cite{F2}, \cite{F3}), larger quantum devices are becoming increasingly attainable; as the complexity and size of these devices grow, hand optimizations will become less feasible, and instead the need for automated design tools will increase. Clearly, it will be highly desirable to have efficient tools for automated circuit optimization, as well as economic, pre-computed circuits for common operations.

In this paper, we present an algorithm for computing an optimal circuit implementing a given unitary transformation on $n$ qubits, with a roughly square root speedup over the brute force algorithm. While the algorithm is tuned to find circuits optimal in terms of depth, it can be adapted to other minimization criteria as well -- we include one such modification to optimize the number of sequential non-Clifford gates in a circuit. We do however note that as the brute force algorithm has exponential complexity, ours does as well, and thus the algorithm's usefulness is limited to factoring small circuits.

Over the years much work has been put into synthesizing optimal circuits for classical, specifically, reversible Boolean functions. Shende {\it et al.} \cite{MARKOV} considered synthesis of 3-bit reversible logic circuits using NOT, CNOT, and Toffoli gates, by generating circuit libraries, then iteratively searching through them. More recently, Golubitsky and Maslov \cite{M1} considered the synthesis of optimal 4-bit reversible circuits composed with NOT, CNOT, Toffoli, and the 4-bit Toffoli gates. In their paper, they describe very efficient performance. While the speed of circuit look-ups described in \cite{M1} is not matched in this work, we contend that synthesis of unitary circuits is a more computationally intensive process than reversible circuits, making direct comparison difficult. Nevertheless, we adapt many of the search techniques described in \cite{M1} to attain fast performance.

Hung {\it et al.} \cite{HUNG} considered a problem somewhat closer to that of quantum circuit synthesis. They developed a method for computing optimal cost decompositions of reversible logic into NOT, CNOT, and the quantum controlled-$\sqrt{X}$ gate. By applying techniques from formal verification, they found minimum cost quantum circuit implementations of various logic gates, including the Toffoli, Fredkin, and Peres gates. However, they use a restricted quantum circuit model (for example, controls are required to remain Boolean), and only a finite subset of quantum circuits on $n$ qubits can be described using four valued logic. Our work by comparison optimizes over all quantum circuits (infinitely many of them) using any gate set, while at the same time generating circuits for all quantum gates, not just Boolean ones, and allowing optimization over other cost functions. Maslov and Miller \cite{M2} also examined synthesis of 3-bit circuits over this gate set, using a pruned breadth-first search similar to \cite{MARKOV} rather than formal methods.

The problem of optimal quantum circuit synthesis is much less studied, and most of the existing work has been focused on finding approximations in small state spaces -- while in this work we focus on finding exact decompositions for various logical gates, the algorithm may be extended to find approximation gate sequences. Dawson and Nielsen \cite{D1} provide an algorithm for computing $\epsilon$-approximations of 1-qubit gates in time $O(\log^{2.71}(1/\epsilon))$, along with a generalization to multi-qubit cases. Their algorithm, offering a constructive proof of the Solovay-Kitaev theorem \cite{SK}, quickly finds a logarithmic (in the precision) depth $\epsilon$-approximation for a given unitary, though the circuit produced may be far from optimal \cite{BS, V1}. It proceeds by recursively computing approximations of unitaries, which in the base case reduces to searching through sets of previously generated unitaries for a basic approximation. By comparison, our algorithm finds a minimal depth circuit, and could in fact also be used to speed up the base case searching in the Solovay-Kitaev algorithm. 

Perhaps more closely related to our work, Fowler \cite{F1} describes an exponential time algorithm for finding depth-optimal $\epsilon$-approximations of single qubit gates. His algorithm uses previously computed knowledge of equivalent subsequences to remove entire sequences from consideration. We believe our algorithm, however, provides better asymptotic behaviour, and that our methods of reducing the search space are in fact more effective. 

More recently, Bocharov and Svore \cite{BS} developed a depth-optimal canonical form for single qubit circuits over the gate set $\{ H, T\}$. As a consequence of their canonical form, they provide a speed up over brute force searching that finds depth-optimal $\epsilon$-approximations by searching through databases of canonical circuits. In this respect our work is somewhat similar, though their canonical form applies only to single qubit circuits over $\{H, T\}$, while our method applies to $n$-qubit circuits over any gate set, and requires significantly less RAM at the expense of slower searches.  Furthermore, the focus of this paper is on the synthesis of small optimal {\em multi-qubit} circuits, that optimal {\em single} qubit synthesis algorithms such as \cite{BS, V1} are unable to tackle. 

The paper is organized as follows: Section 2 describes the mathematical background and framework upon which we build our algorithm; Section 3 gives a description of our algorithm; Section 4 discusses methods of reducing the search space; Section 5 describes details of our implementation; Section 6 presents results of our computations and performance figures, and Section 7 concludes the paper and discusses future work.

\section{Preliminaries}

In this section we define some of the mathematical ideas and notation used throughout the paper. We will assume the reader is familiar with the basics of quantum computation, but review the quantum circuit model for completeness.

In the circuit model of quantum computation, {\it wires} carry quantum bits ({\it qubits}) to gates, which transform their state. Given that the state of a system of $n$ qubits is typically described as a vector in a $2^n$-dimensional complex vector space $\mathcal{H}$, quantum gates can be viewed as linear operators on $\mathcal{H}$. We restrict attention to unitary operators, i.e. operators $U$ such that $UU^\dagger=U^\dagger U=I$, where $U^\dagger$ denotes the adjoint (conjugate-transpose) of $U$ and $I$ denotes the identity operator. The linear operator performed by the circuit is then given as the sequential composition of the individual gates within the circuit, and it is easily verified that this linear operator is itself unitary.

An individual gate acts non-trivially only on a subset of the qubits in a given system. As such, it is convenient to describe the unitary performed by the gate as the tensor product of the non-trivial unitary on a smaller state space corresponding to the affected qubits, and the identity operator on the remaining qubits. This presentation of gates also displays the parallel nature of circuits: a sequential circuit composed of two gates $g_1, g_2$ represented by unitaries $(g_1\otimes I)$ and $(I\otimes g_2)$ can be rewritten in parallel as $g_1\otimes g_2$.

As our main optimization criteria for circuits, we define the {\it depth} of a circuit as the length of any critical path through the circuit. Representing a circuit as a directed acyclic graph with nodes corresponding to the circuit's gates and edges corresponding to gate inputs/outputs, a critical path is a path of maximum length flowing from an input of the circuit to an output.

\begin{figure}[!t]
\centering
\includegraphics[scale=0.45]{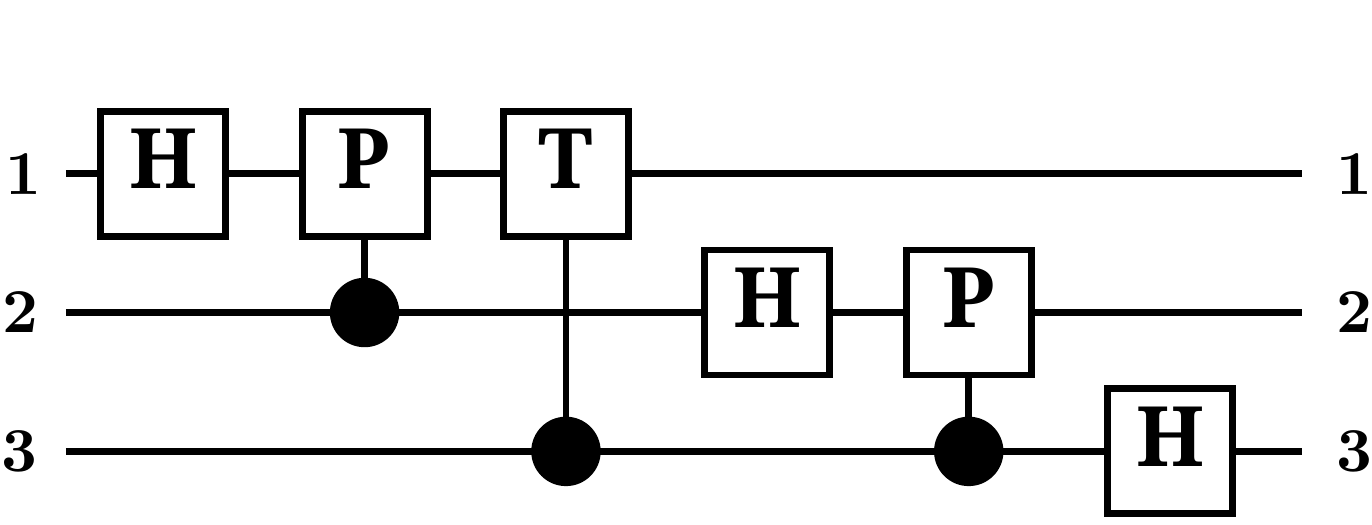}
\caption{A quantum circuit performing the quantum Fourier transform, up to permutation of the outputs. This circuit has depth 5, with two critical paths flowing from input 1 to output 3.}
\label{fig:QFT}
\end{figure}

The problem of quantum circuit synthesis refers to finding some circuit containing only gates taken from a fixed set performing the desired unitary. We call this fixed set an {\it instruction set}, and require that it contains the inverse of each gate in the set. An $n$-qubit circuit over instruction set $\mathcal{G}$ is then the composition of individual gates applied to non-empty subsets of $n$-qubits, tensored with the identity on the remaining qubits. 

We may combine gates from an instruction set, each acting on different qubits, to construct circuits of depth one over $n$ qubits. As such circuits will be integral to our algorithm, we define $\mathcal{V}_{n,\mathcal{G}}$, the set of all unitaries corresponding to depth one $n$-qubit circuits over the instruction set $\mathcal{G}$. An $n$ qubit circuit $C$ over the instruction set $\mathcal{G}$ then has depth at most $m$ if $C$ corresponds to some sequence of unitaries $U_1U_2\cdots U_m$ where $U_1,U_2,...,U_m\in\mathcal{V}_{n,\mathcal{G}}$. Additionally, we say that $C$ implements a unitary $U\in U(2^n)$ if $U_1U_2\cdots U_m=U$.

In general, many distinct circuits may implement the same unitary. While we frequently will not distinguish a circuit from the unitary it implements, it is assumed that by ``circuit" we mean some specific sequence of gates, rather than the resulting unitary transformation. Also important to note is that circuits are written in terms of operator composition, so unitaries are applied right to left -- in a circuit diagram, however, gates are applied left to right.

With these definitions, we are now able to formulate our main result. In particular, we present an algorithm that, given an instruction set $\mathcal{G}$ and unitary transformation $U\in U(2^n)$, determines whether $U$ can be implemented by a circuit over $\mathcal{G}$ of depth at most $l$ in time\footnote{We assume the RAM model of computation throughout this paper.} $O\big(|\mathcal{V}_{n,\mathcal{G}}|^{\lceil l/2 \rceil}\log(|\mathcal{V}_{n,\mathcal{G}}|^{\lceil l/2 \rceil})\big)$. Furthermore, if $U$ can be implemented with a circuit of depth at most $l$, the algorithm returns a circuit implementing $U$ in minimal depth over $\mathcal{G}$. We also present our C/C++ implementation of the algorithm and report better circuits, generated with our implementation, than those found in the literature.

This algorithm provides a clear improvement over the brute force $O(|\mathcal{V}_{n,\mathcal{G}}|^l)$ algorithm. In practice, it is possible to achieve running times close to $\Theta(|\mathcal{V}_{n,\mathcal{G}}|^{\lceil l/2 \rceil})$ by careful construction of data structures. However, we stress that despite the roughly quadratic speed-up, the runtime is still exponential, since $|\mathcal{V}_{n,\mathcal{G}}|\geq k^n$ for any instruction set $\mathcal{G}$ with $k$ single qubit gates, making it only practical for small numbers of qubits.

Motivated by results in fault tolerance, we use the instruction set consisting of the Hadamard gate $H=\frac{1}{\sqrt{2}}\begin{pmatrix} 1 & 1 \\ 1 & -1 \end{pmatrix}$, Phase gate $P=\begin{pmatrix} 1 & 0 \\ 0 & i \end{pmatrix} $, controlled-NOT $CNOT= \begin{pmatrix} 1 & 0 & 0 & 0 \\ 0 & 1 & 0 & 0 \\ 0 & 0 & 0 & 1 \\ 0 & 0 & 1 & 0 \end{pmatrix},$ and $T=\begin{pmatrix} 1 & 0 \\ 0 & e^{\frac{i\pi}{4}} \end{pmatrix}$, along with $P^\dagger$ and $T^\dagger$. The set of circuits composed from these gates forms a subset of $2^n\times2^n$ unitary matrices over the ring $\mathbb{Z}\left[\frac{1}{\sqrt{2}}, i\right]$ defined as $$\mathbb{Z}\left[\frac{1}{\sqrt{2}}, i\right]=\left\{\left. \frac{a + be^{i\frac{\pi}{4}} + ce^{i\frac{\pi}{2}} + de^{i\frac{3\pi}{4}}}{\sqrt{2^n}} \right| \text{\parbox{0.89in}{\centering $a, b, c, d, n\in\mathbb{Z}$, \\ $n\geq 0$}}\right\}.$$ We also identify two important groups of matrices over this ring: the Pauli group on $n$-qubits, denoted $\mathcal{P}_n$ and defined as the set of all $n$-fold tensor products of the Pauli matrices $I=\begin{pmatrix} 1 & 0 \\ 0 & 1 \end{pmatrix}, X=\begin{pmatrix} 0 & 1 \\ 1 & 0 \end{pmatrix}, Y=\begin{pmatrix} 0 & -i \\ i & 0 \end{pmatrix}, Z=\begin{pmatrix} 1 & 0 \\ 0 & -1 \end{pmatrix}$, and also the Clifford group on $n$-qubits, denoted $\mathcal{C}_n$ and defined as the normalizer\footnote{$\mathcal{C}_n=\{U\in U(2^n)|U\mathcal{P}_nU^{-1}\subseteq\mathcal{P}_n\}$.} of $\mathcal{P}_n$ in $U(2^n)$. In particular, unitaries computable by circuits composed with $I, X, Y, Z, H, P$, and $CNOT$ are elements of the Clifford group, while the $T$ gate does not belong to the Clifford group.

As a well-known result, $\mathcal{C}_n$ along with any one unitary $U\notin \mathcal{C}_n$ forms a set dense in $U(2^n)$ \cite{SK}. Since $\{H, P, CNOT\}$ generates the Clifford group up to global phase, the instruction set consisting of $\{H, P, P^\dagger, CNOT, T, T^\dagger\}$ is universal for quantum computing.

\section{Search Algorithm}

This section gives a high level description of the algorithm we use to compute optimal circuits; Section 5 will go into more depth regarding the implementation. Also, we only describe the case in which we are searching for circuits that implement a given unitary exactly -- discussion on extending the algorithm to find approximating circuits will be left to Section 7.

The main insight of our algorithm is the following observation, which allows us to search for circuits of depth $l$ by only generating circuits of depth at most $\lceil l/2 \rceil$.

\begin{lemma}\label{lem:optimal}
Let $S_i\subset U(2^n)$ be the set of all unitaries implementable in depth $i$ over the gate set $\mathcal{G}$.
Given a unitary $U$, there exists a circuit over $\mathcal{G}$ of depth $l$ implementing $U$ if and only if $S_{\lfloor l/2\rfloor}^\dagger U\cap S_{\lceil l/2\rceil}\neq\emptyset$.
\end{lemma}
\noindent
{\em Proof:}
Before proving the lemma, we see that $U\in S_i^\dagger = \{ U^\dagger | U\in S_i\}$ if and only if $U$ can be implemented in depth $i$ over $\mathcal{G}$. In particular, for any unitary $U=U_1U_2\cdots U_i$ where $U_1,U_2,\dots ,U_i\in\mathcal{V}_{n, \mathcal{G}}$, we see that $U^\dagger = (U_1U_2\cdots U_i)^\dagger = U_i^\dagger\cdots U_2^\dagger U_1^\dagger$ as a basic result of linear algebra. As $\mathcal{G}$ is closed under inversion, $U_1^\dagger,U_2^\dagger,\dots ,U_i^\dagger\in\mathcal{V}_{n, \mathcal{G}}$, and thus a circuit of depth $i$ over $\mathcal{G}$ implements $U^\dagger$. The reverse direction can be observed by noting that $(S_i^\dagger)^\dagger=S_i$.

Now we prove the lemma. Suppose some depth $l$ circuit $C$ implements $U$. We divide $C$ into two circuits of depth $\lfloor l/2 \rfloor$ and $\lceil l/2 \rceil$, implementing unitaries $V\in S_{\lfloor l/2\rfloor}$ and $W\in S_{\lceil l/2\rceil}$ respectively, where $VW=U$. Since we know $W=V^\dagger U\in S_{\lfloor l/2\rfloor}^\dagger U$, we can observe that $W\in S_{\lfloor l/2\rfloor}^\dagger U\cap S_{\lceil l/2\rceil}$, as required.

Suppose instead $S_{\lfloor l/2\rfloor}^\dagger U\cap S_{\lceil l/2\rceil}\neq\emptyset$. We see that there exists some $W\in S_{\lfloor l/2\rfloor}^\dagger U\cap S_{\lceil l/2\rceil}$, and moreover by definition $W=V^\dagger U$ for some $V^\dagger\in S_{\lfloor l/2\rfloor}^\dagger$. Since $W\in S_{\lceil l/2\rceil}$, $VW=U$ is implementable by some circuit of depth $\lfloor l/2\rfloor + \lceil l/2\rceil=l/2$, thus completing the proof.
\hfill $\Box$

We use this lemma to develop a simple algorithm to determine whether there exists a circuit over $\mathcal{G}$ of depth at most $l$ implementing unitary $U$, and if so return a minimum depth circuit implementing $U$.

\vspace{0.3cm}
\hrule
\vspace{0.1cm}
\begin{algorithmic}
\Function{MITM-factor}{$\mathcal{G}$, $U$, $l$} \\
	\State $S_0 := \{I\}$
	\State $i := 1$
	\For{$i\leq \lceil l/2\rceil$}
		\State $S_i := \mathcal{V}_{n,\mathcal{G}} S_{i-1}$
		\If{$S_{i-1}^\dagger U\cap S_i\neq\emptyset$}
			\State \Return{any circuit $VW$ s.t.}
			\State \qquad $V\in S_{i-1}, W\in S_i, V^\dagger U=W$
		\ElsIf{$S_{i}^\dagger U\cap S_i\neq\emptyset$}
			\State \Return{any circuit $VW$ s.t.}
			\State \qquad $V, W\in S_i, V^\dagger U=W$
		\EndIf
		\State $i := i+1$
	\EndFor
\EndFunction
\end{algorithmic}
\vspace{0.1cm}
\hrule
\vspace{0.3cm}

\begin{figure}[!t]
\centering
\includegraphics[scale=0.3]{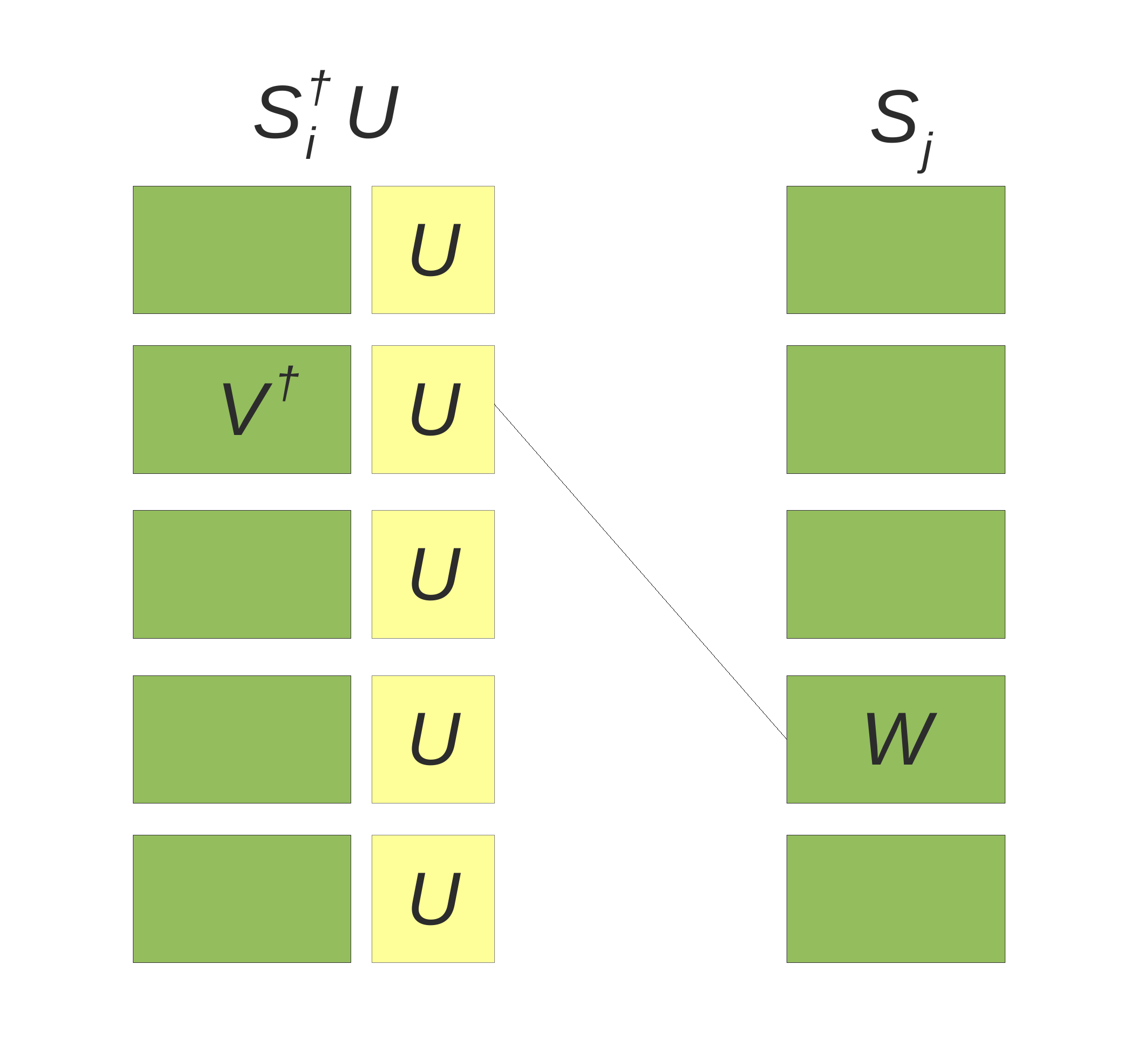}
\caption{For each $V\in S_i$ we construct $W=V^\dagger U$ and perform a logarithmic-time search for $W$ in $S_j$.}
\label{fig:mitm}
\end{figure}

Given an instruction set $\mathcal{G}$ and unitary $U$, we repeatedly generate circuits of increasing depth, then use them to search for circuits implementing $U$ with up to twice the depth (Figure~\ref{fig:mitm}). Specifically, at each step we generate all depth $i$ circuits $S_i$ by extending the depth $i-1$ circuits with one more level of depth, then we compute the sets $S_{i-1}^\dagger U$ and $S_{i}^\dagger U$ and see if there are any collisions with $S_i$. By Lemma \ref{lem:optimal}, there exists a circuit of depth $2i-1$ or $2i$ implementing $U$ if and only if $S_{i-1}^\dagger U\cap S_i\neq\emptyset$ or $S_{i}^\dagger U\cap S_i\neq\emptyset$, respectively, so the algorithm terminates at the smallest depth less than or equal to $l$ for which there exists a circuit implementing $U$. In the case where $U$ can be implemented in depth at most $l$, the algorithm returns one such circuit of minimal depth.

To observe the claimed $O\big(|\mathcal{V}_{n,\mathcal{G}}|^{\lceil l/2 \rceil}\log(|\mathcal{V}_{n,\mathcal{G}}|^{\lceil l/2 \rceil})\big)$ runtime, we impose a strict lexicographic ordering on unitaries -- as a simple example two unitary matrices can be ordered according to the first element on which they differ. The set $S_i$ can then be sorted with respect to this ordering in $O\big(|S_i|\log(|S_i|)\big)$ time, so that searching for each element of $S_{i-1}^\dagger U$ and $S_i^\dagger U$ in $S_i$ can be performed in time $O\big(|S_{i-1}|\log(|S_i|)\big)$ and $O\big(|S_i|\log(|S_i|)\big)$, respectively.  As $|S_i|\leq|\mathcal{V}_{n,\mathcal{G}}|^i$, the $i$th iteration thus takes time bounded above by $2|\mathcal{V}_{n,\mathcal{G}}|^i\log(|\mathcal{V}_{n,\mathcal{G}}|^i)$. Since $\sum_{i=1}^{\lceil l/2\rceil}|\mathcal{V}_{n,\mathcal{G}}|^i\log(|\mathcal{V}_{n,\mathcal{G}}|^i)\leq\sum_{i=1}^{\lceil l/2\rceil}|\mathcal{V}_{n,\mathcal{G}}|^i\log(|\mathcal{V}_{n,\mathcal{G}}|^{\lceil l/2\rceil})$ and  $\sum_{i=1}^{\lceil l/2\rceil}|\mathcal{V}_{n,\mathcal{G}}|^i\leq |\mathcal{V}_{n, \mathcal{G}}|^{\lceil l/2 \rceil}\big(1+\frac{1}{|\mathcal{V}_{n, \mathcal{G}}|^{\lceil l/2 \rceil} - 1}\big)$, we thus see that the algorithm runs in $O\big(|\mathcal{V}_{n,\mathcal{G}}|^{\lceil l/2 \rceil}\log(|\mathcal{V}_{n,\mathcal{G}}|^{\lceil l/2 \rceil})\big)$ time. 

It can also be noted that $\mathcal{V}_{n,\mathcal{G}}\in O(|\mathcal{G}|^n)$, so the runtime is in $O\big(|\mathcal{G}|^{\lceil n\cdot l/2 \rceil}\log(|\mathcal{G}|^{\lceil n\cdot l/2 \rceil})\big)$.

\subsection{Optimizing different cost functions}

This algorithm also allows one to search for circuits where optimality is given by some other criteria, though the runtime will still be parameterized on the depth of the solution. For example, we can consider cases where circuit cost is defined as a weighted gate sum, with weights assigned to the instruction set elements. As long as the weights are strictly positive on all (non-identity) gates, the minimum circuit cost will be a strictly increasing function of circuit depth, so the cost of any solution provides an upper bound on the depth that needs to be explored.

We focus, however, on one specific family of cost functions that is becoming increasingly important in fault tolerant models. In many of the common models the Clifford group has an efficient set of generators, while non-Clifford group gates require more expensive procedures to implement. For fault tolerant quantum computing based on the Steane code, non-Clifford group gates are significantly more complicated \cite{G1}; in fact, for all double even and self-dual CSS codes, a class that includes many common quantum error correcting codes, all Clifford group operations have transversal implementations \cite{L1} and thus are relatively simple to implement. By contrast, non-Clifford gates require much more sophisticated and costly techniques to implement, such as ancilla preparation and gate teleportation. The more recent surface codes, which promise higher thresholds than concatenated code schemes, also have a significantly more complicated $T$-gate implementation than any of the Clifford group generators \cite{F2}. As a result, the number of stages in the circuit involving non-Clifford group gates -- called the circuit's {\it $T$-depth} when the $T$ gate is the only non-Clifford gate -- becomes the bottleneck in fault-tolerant computation.

Fortunately, our algorithm permits an easy modification to optimize circuit $T$-depth. In cases where the instruction set $\mathcal{G}$ is given as a set of generators for the Clifford group as well as the $T$ gate, we can generate the entire Clifford group then by brute force enumerate all circuits in increasing $T$-depth. Then, using the meet-in-the-middle technique, circuits with up to twice the $T$-depth can be searched for a match.

Specifically, we define $\mathcal{C}_n$ to be the Clifford group on $n$ qubits, as implemented by the gates in $\mathcal{G}$,  and $\mathcal{T}_n$ to be the set of tensor products of $I$ and $T$. To perform the meet-in-the-middle search, we set $S_0=\mathcal{C}_n$, and $S_i=\mathcal{C}_n\left(\mathcal{T}_n\setminus \{I\}\right)S_{i-1}$; each $S_i$ thus contains every circuit with $T$-depth $i$. Searching then proceeds by computing the intersections $S_{i-1}^\dagger U\cap S_i$ to search for $T$-depth $2i-1$ circuits, and $S_{i}^\dagger U\cap S_i$ for $T$-depth $2i$. The full algorithm is summed up in pseudocode below.

\vspace{0.3cm}
\hrule
\vspace{0.1cm}
\begin{algorithmic}
\Function{MITM-factor $T$-depth}{$\mathcal{G}$, $U$, $l$} \\
	\State $S_0 := \{\mathcal{C}_n\}$
	\State $i := 1$
	\For{$i\leq l$}
		\State $S_i := \left(\mathcal{C}_n\mathcal{T}_n\setminus\{I\}\right)S_{i-1}$
		\If{$S_{i-1}^\dagger U\cap S_{i}\neq\emptyset$}
			\State \Return{any circuit $VW$ s.t.}
			\State \qquad $V\in S_{i-1}, W\in S_i, V^\dagger U=W$
		\EndIf
		\If{$S_{i}^\dagger U\cap S_{i}\neq\emptyset$}
			\State \Return{any circuit $VW$ s.t.}
			\State \qquad $V\in S_{i}, W\in S_i, V^\dagger U=W$
		\EndIf
		\State $i := i+1$
	\EndFor
\EndFunction
\end{algorithmic}
\vspace{0.1cm}
\hrule
\vspace{0.3cm}

Searching in this way however becomes challenging for high dimensional state spaces, as the size of the Clifford group grows exponentially in the number of qubits. As an illustration, for 3 qubits the Clifford group has 92,897,280 elements up to global phase \cite{G2}, which makes searching using modern computers impractical for more than a couple levels of depth; $\mathcal{C}_4$ would not even fit in a computer with a reasonable amount of memory using this method. 

In practice we compute sets $S_i$ with $T$-depth $\lceil i/2 \rceil$ instead, by alternating between Clifford and $T$ phases. This allows a large amount of redundancy in the meet in the middle computation to be removed, as an entire phase of Clifford group gates can be ignored when searching. Given the enormous size of the Clifford group, this provides serious performance advantages over the more na\"{i}ve algorithm shown above.

\subsection{Circuits with ancillas}

As another useful extension of our algorithm, it can be employed to search for circuits that make use of ancillary qubits, initialized to $|0\rangle$ and returned to the state $|0\rangle$. In general, it may be possible to consider arbitrary ancilla states, but for simplicity we only allow the zero state, and by extension any state that can be prepared with the instruction set.

Specifically, if $U\in U(2^n)$ and we want to find some $U'\in U(2^{n+m})$ so that $U'(|0\rangle^{\otimes m}|\psi\rangle)=|0\rangle^{\otimes m}(U|\psi\rangle)$, we need only look for some $U'$ that agrees with $U$ on the first $2^n$ rows and columns. However, if we only use the first $2^n$ rows and columns of each unitary to perform searching, many collisions may be lost, as it may be the case that $V'W'(|0\rangle^{\otimes m}|\psi\rangle)=|0\rangle^{\otimes m}(U|\psi\rangle)$ but $V'(|0\rangle^{\otimes m}|\psi\rangle)\neq|0\rangle^{\otimes m}(V|\psi\rangle)$, and likewise for $W$.

Instead, we note that since $|0\rangle^{\otimes m}(U|\psi\rangle)=(I\otimes U)|0\rangle^{\otimes m}|\psi\rangle$, we want to find $V, W$ such that $VW(|0\rangle^{\otimes m}|\psi\rangle)=(I\otimes U)|0\rangle^{\otimes m}|\psi\rangle$. Then, clearly $V^\dagger(I\otimes U)|0\rangle^{\otimes m}|\psi\rangle=W(|0\rangle^{\otimes m}|\psi\rangle)$, so we can restrict the sets $S_{\{i-1, i\}}^\dagger(I\otimes U)$ and $S_{i}$ to inputs of the form $|0\rangle^{\otimes m}|\psi\rangle$ and determine whether their intersection is non-zero. Practically speaking, this involves only comparing the first $2^n$ columns of each possible collision -- however, more circuits need to be generated and searched since circuit permutations can no longer be removed, as described for the main algorithm in the following section.

\section{Search tree pruning}
To further reduce the search space, we prune the search tree\footnote{By search tree we mean the tree where each branch corresponds to a different choice of the next gate (from $\mathcal{V}_{n, \mathcal{G}}$) in the circuit. Each $S_i$ corresponds to one level of depth in the tree, and our algorithm then generates the tree breadth first.}, in practice providing significant reductions in both space and time used.

To prune the search tree, we define an equivalence relation $\sim$ on unitary transformations where $U\sim V$ if and only if $U$ is equal to $V$ up to relabeling of the qubits, inversion, or global phase factors. This equivalence relation defines the equivalence class of a unitary $U\in U(2^n)$, denoted $[U]$, as $\{ V\in U(2^n) | U\sim V\}$. We then store only one minimal depth circuit implementing the representative of each unitary equivalence class; specifically, we define a canonical representative for each unitary equivalence class, then when a new circuit is generated we find the unitary representative and determine whether a circuit implementing it is already known. 

We define a canonical representative for each unitary equivalence class by lexicographically ordering unitaries and choosing the smallest unitary as the representative. Since relabeling of the qubits corresponds to simultaneous row and column permutations of the unitary matrix and the inverse of a unitary is given by its conjugate transpose, given an $n$ qubit unitary $U$, all $2n!$ permutations and inversions of $U$ can be generated and the minimum can be found in $O(n!)$ time. The added $O(n!)$ overhead per unitary has little effect on the overall run-time for small $n$, as the time to compute a canonical unitary is minimal compared to the time to search for a unitary in each circuit database.

Choosing a canonical unitary up to phase is more difficult in general. In the case when the instruction set can be written as unitaries over the ring $\mathbb{Z}\left[\frac{1}{\sqrt{2}}, i\right]$, we can generate each possible global phase factor in the equivalence class. In particular, $e^{i\theta}\in \mathbb{Z}\left[\frac{1}{\sqrt{2}}, i\right]$ if and only if $\theta=\frac{k\pi}{4}, k\in \mathbb{Z}$ \cite{V1}, so there are only 8 possible global phase factors for any unitary over $\mathbb{Z}\left[\frac{1}{\sqrt{2}}, i\right]$. To find a representative, all $8\cdot 2n!$ elements of $[U]$ are generated, and only the lexicographically earliest element is kept.

In practice, computing each phase factor causes a significant performance hit, so we sought a more efficient way of removing phase equivalences. Instead, we pick a reference element for each unitary and use it to define the canonical phase of the unitary -- by convention we choose the first (scanning row by row) non-zero element of a unitary matrix. For a reference element $re^{i\theta}$ of a unitary $U$, we can define the canonical unitary as $e^{-i\theta}U$, so that if $V=e^{i\phi}U$, the reference of $V$ will be $re^{i(\theta+\phi)}$ and so $e^{-i(\theta+\phi)}V=e^{-i\theta}U$. If $\theta\neq\frac{k\pi}{4}, k\in\mathbb{Z}$, this phase multiple will take $U$ outside of the ring $\mathbb{Z}\left[\frac{1}{\sqrt{2}}, i\right]$, so we actually remove the constraint that the canonical unitary is normalized instead; rather than taking $e^{-i\theta}U$ as the canonical unitary, we use $re^{-i\theta}U$ since $re^{-i\theta}\in\mathbb{Z}\left[\frac{1}{\sqrt{2}}, i\right]$. While this is a minor detail, it allows comparisons to be performed symbolically over the ring, allowing much more accurate and expedient computations.

Given a newly generated circuit $C$ with depth $i$ implementing unitary $U$, the canonical representative of $[U]$ is computed then the database of circuits is searched to determine if another circuit implementing the representative has already been found. Using suitable data structures, each previous set $S_j$, $1\leq j\leq i$ can be searched in $O\big(\log(|S_j|)\big)$ time. If no such circuit is found, we store a circuit implementing the representative of $[U]$. It suffices to observe that given a circuit $C$ implementing $U$, a circuit implementing the representative can be computed by applying the corresponding permutation and/or inversion to $C$. Permutations are applied by changing which qubit the individual gates in the circuit act on; for inverses, we note that $C^{-1}=C^\dagger$, so $C=U_1\cdots U_m$ implies $C^{-1} = U_m^\dagger\cdots U_1^\dagger$. Since our definition of an instruction set required that each gate have an inverse in the set, a circuit for $C^{-1}$ is obtained by reversing the gates of $C$ and replacing them with their own inverse. As a consequence, each permutation and inverse of a unitary can be implemented in the same depth, so every unitary in an equivalence class has the same minimum circuit depth.

As a subtle point, if only representatives of equivalence classes in depth $i$ and $j$ are used for searching, not every equivalence class in depth $i+j$ will be found. Consider some unitary $U=VW$ where $V$ is a circuit of depth $i$, and $W$ is a circuit of depth $j$. If $V'$ is the representative of $[V]$ and $W'$ is the representative of $[W]$, then in general $(V')^\dagger VW\notin [W]$, so just using class representatives to search will not suffice. However, $V^\dagger\in[V']$, so $W\in[V']VW$, and thus $[V']U=[W']$. Practically speaking, this means that any unitary $U=VW$ is found by computing the canonical representatives of $[[V]U]$, and so we can search all circuits in minimum depth by storing only equivalence class representatives.

In some cases an exact implementation with the same global phase is required, particularly when the circuit may be controlled on another qubit. While we could compute canonical representatives with respect to qubit relabeling and inversion only, any canonical phase implementation over $\mathcal{G}=\{ H, P, P^\dagger, CNOT, T, T^\dagger\}$ can be used to construct the correct global phase, if it is implementable over $\mathcal{G}$. It suffices to observe (as follows from \cite{V1}) that if $U$ is implementable by a circuit over $\mathcal{G}$ and a circuit $C$ over $\mathcal{G}$ implements $e^{i\theta}U$, then $\theta=\frac{k\pi}{4}$ for some $k\in \mathbb{Z}$. Since $(HP^\dagger)^3=e^{-i\frac{\pi}{4}}I$, $e^{i\theta}(HP^\dagger)^{3k}U=U$, so a circuit implementing $U$ exactly can be generated using $C$.

\section{Implementation Details}

It was mentioned earlier that the meet-in-the-middle algorithm offers no speed up over the na\"{i}ve algorithm without suitably chosen data structures, as searching for collisions in unordered sets $S_i^\dagger U$ and $S_j$ would use $O(|S_i||S_j|)$ comparisons. However, by imposing a lexicographic ordering on the generated circuits, they can be searched in time logarithmic in the size of $S_j$. In our implementation, we use such an ordering to store each $S_i$ as a red-black tree, a type of balanced binary tree. Balanced binary trees are a common choice for implementations of ordered sets and mappings in standard libraries, as well as industrial databases, due to predictable performance and scalability. Since deletions, typically the most computationally difficult task in a balanced binary tree \cite{T1}, are never performed in our implementation, such trees are a natural choice of data structure. 

While hash tables could potentially provide a speed up, our code was adapted to use the hash table implementation in libstd++ and no performance improvement was found despite very reasonable numbers of key collisions -- as an illustration, at most 52 out of 1,316,882 distinct 3-qubit unitaries were mapped to any one hash value.

While storing the unitaries themselves allows fast generation of new unitaries, as well as fast searching through circuit databases, the storage space required makes large searches impossible on computers with reasonable sized RAM. Given an instruction set $\mathcal{G}$ over $n$ qubits, $|\mathcal{V}_{n,\mathcal{G}}|\geq k^n$ where $k$ denotes the number of one qubit gates in $\mathcal{G}$. Using the standard universal instruction set, $\{H, P, P^\dagger, CNOT, T, T^\dagger\}$ (i.e. a generating set for the Clifford group plus the $T$ gate), $|\mathcal{V}_{3,\mathcal{G}}|=252$, so even at depth 5 there are more than $10^{12}$ circuits. If unitaries over $\mathbb{Z}\left[\frac{1}{\sqrt{2}}, i\right]$ are stored exactly, each $3$-qubit unitary requires $5\times 64$ integers, and so all depth 5 circuits on 3 qubits would require more than $1$ petabyte of storage space. In reality storing only equivalence class representatives makes a significant difference (for 3 qubits, there are at most 36,042,958 unique equivalence classes with circuits up to depth 5 according to our experiments), and the storage space for unitaries over $\mathbb{Z}\left[\frac{1}{\sqrt{2}}, i\right]$ could be reduced by applying compression; yet, it is still clear that for searches up to significant depth, the full unitary matrices cannot be stored and a space-time trade off must be made.

To make such a trade off, rather than store the generated unitaries, we store the circuit as a list of depth 1 circuits. Each depth 1 circuit on $n$-qubits is represented as $n$ bytes specifying which gate is acting on each qubit. However, if only the circuit was stored searching for a specific unitary would take an excessive amount of time, since each time a comparison is invoked the unitary implemented by the circuit would need to be computed again. 

As a compromise, an $m\times m$ matrix $M$ is stored as a key with each circuit, where for a circuit $C$ implementing unitary $U$, $M(i, j)=v_i^\dagger Uv_j$. The $m$ vectors $\{v_i\}$ are chosen from $\mathbb{C}^{2^n}$ using a pseudorandom generator to generate the individual elements, and in practice $m=1$ has been enough to search interesting depths for up to 4 qubits with extremely few key collisions. Since these keys are generated with floating point computations, it's important that all other computations are performed symbolically and keys are computed directly from the unitary, so that equal unitaries will have equal numerical error. Experiments were also performed using random vectors over $\mathbb{Z}\left[\frac{1}{\sqrt{2}}, i\right]$ to avoid all floating point computations, but they generated far too many key collisions to be of practical use. Currently we are looking into better methods for generating unitary keys.

Additionally, to improve performance for circuit searching, our code is parallelized (using the pthreads C library) to allow searches in balanced binary trees to be computed concurrently on different threads. Generated circuit databases are also serialized and stored in files so they are not re-computed for every search.

\section{Performance and Results}\label{sec:performance}

We tested our implementation in Debian Linux on our group's research server, containing a quad-core, 64-bit Intel Core i5 2.80GHZ processor and 16 GB RAM, plus an additional 16 GB of swap space.

\begin{figure}[h]
\centering
\includegraphics[scale=0.80]{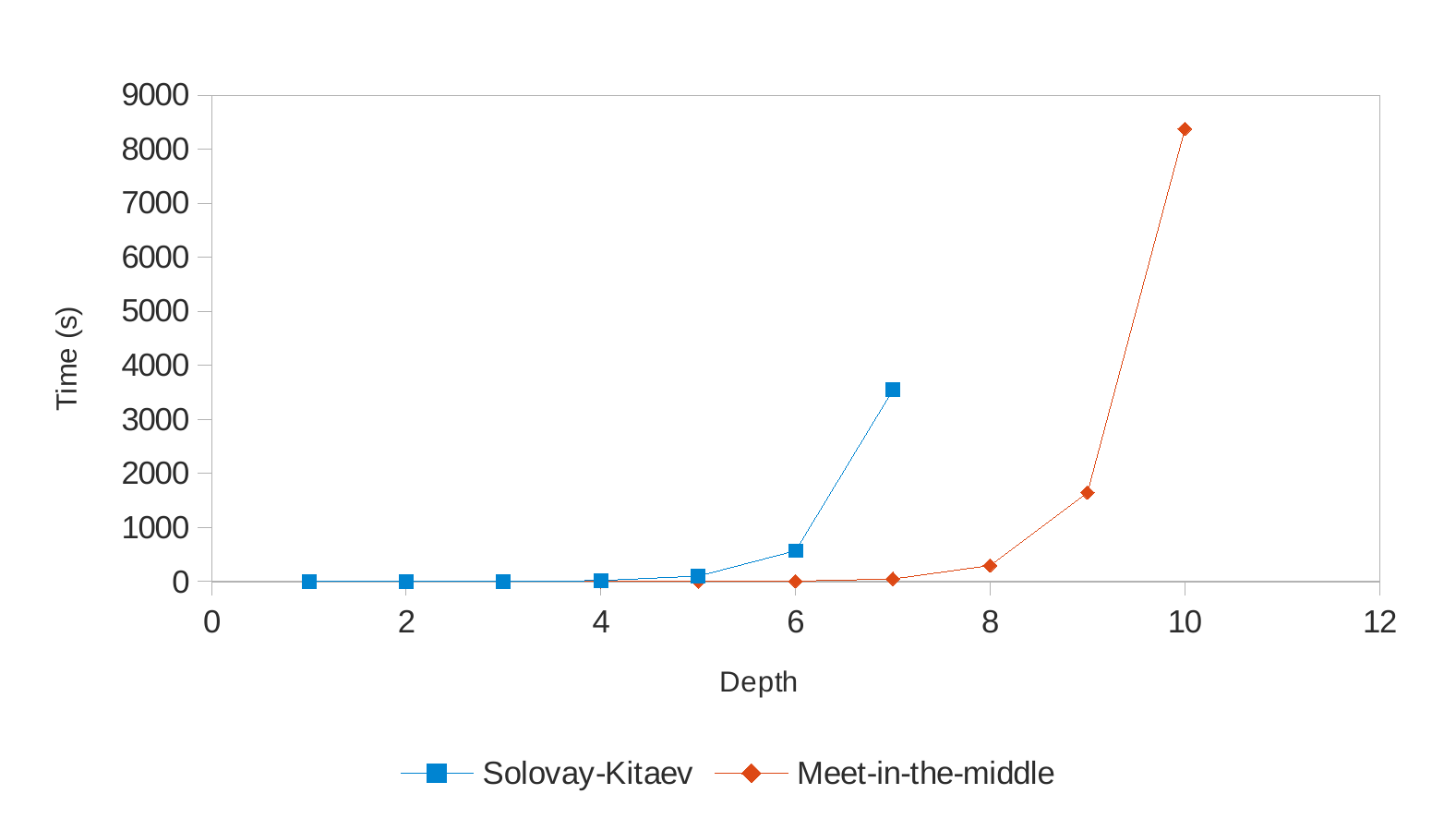}
\caption{Database generation times for minimal depth two qubit circuits.}
\label{fig:sk}
\end{figure}

We compared our implementation of the meet-in-the-middle algorithm with an open-source Python implementation \cite{PAUL} of the Solovay-Kitaev algorithm \cite{SK}. This particular implementation was chosen over faster C versions or exact synthesis tools as it is the only existing tool to our knowledge decomposing multiple qubit operators over the Clifford + $T$ gate set. Database generation times for two qubit circuits composed with $H$, $T$, $T^\dagger$, and $CNOT$ are shown in Figure~\ref{fig:sk}; the meet-in-the-middle implementation shows a similar but compressed exponential curve.

We compared our implementation of the meet-in-the-middle algorithm with an open-source Python implementation \cite{PAUL} of the Solovay-Kitaev algorithm \cite{SK}. This particular implementation was chosen over faster C versions as it is the only tool to our knowledge decomposing multiple qubit operators over the Clifford + $T$ gate set. Database generation times for two qubit circuits composed with $H$, $T$, $T^\dagger$, and $CNOT$ are shown in Figure~\ref{fig:sk}; the meet-in-the-middle implementation shows a similar but compressed exponential curve.

A decomposition of the controlled-$H$ gate was generated with both the Solovay-Kitaev and meet-in-the-middle algorithms. While our algorithm produced an exact decomposition with minimal depth (Figure~\ref{fig:CH}) in under 0.500s, the Solovay-Kitaev algorithm with 4 levels of recursion took over 2 minutes to generate a sequence consisting of over 1000 gates approximating the unitary to an error of 0.340 in Fowler's distance metric \cite{F1}. While we stress that the Solovay-Kitaev algorithm is not designed to factor unitaries exactly over a gate set, this experiment serves to illustrate both the efficiency of our implementation, as well as the limitations of current quantum circuit synthesis tools.

\subsection{Depth-optimal implementations}

We ran experiments on various 2, 3, and 4 qubit logical gates to find optimal depth decompositions into the gate set $\{ H, P, P^\dagger, CNOT, T, T^\dagger\}$, with a secondary optimization criteria being the gate count. Table~\ref{fig:table} lists some performance figures for our implementation -- searching times for a given depth $i$ describe the time the computation took to search through circuits of depth $2i-1$ and $2i$. While searches returning more results run slower, the variance is extremely minor, so one representative search was chosen for each set of data. As one other important point, the search times given are computed by searching for all collisions in $S_j^\dagger U$ and $S_i$ -- if $S_j^\dagger U\cap S_i\neq\emptyset$, collisions are usually found within a few minutes of searching.

\begin{table}[h]
\footnotesize
\renewcommand{\arraystretch}{1.1}
\caption{Performance for depth-optimal circuit searches}
\label{fig:table}
\centering
\begin{tabular}{| c | c | c | c | c | c | c | c |}
\hline
\multicolumn{2}{| l |}{\# qubits $\backslash$ depth} & 1 & 2 & 3 & 4 & 5 & 6 \\ \hline
   & database size (circuits) & 14 & 104 & 901 & 6,180 & 37,878 & 197,388 \\ \cline{2-8}
2 & RAM (KB) & 2.092 & 16.686 & 146.701 & 1,013.358 & 6,249.708 & 32,766.246 \\ \cline{2-8}
   & generation time (s)  & 0.001 & 0.015 & 0.155 & 1.354 & 10.761 & 75.301 \\ \cline{2-8}
   & search time (s)   & 0.001 & 0.004 & 0.033 & 0.248 & 1.672 & 9.321 \\ \hline
   & database size (circuits) & 36 & 1,110 & 41,338 & 1,316,882 & 36,042,958 & - \\ \cline{2-8}
3 & RAM (KB) & 5.633 & 179.657 & 6,737.931 & 215,968.485 & 7,738,582.749 & - \\ \cline{2-8}
   & generation time (s)  & 0.012 & 1.059 & 40.619 & 1896.301 & 73,295.675 & - \\ \cline{2-8}
   & search time (s)   & 0.015 & 0.350 & 12.619 & 414.722 & 11,759.390 & - \\ \hline
   & database size (circuits) & 84 & 9,984 & 1,755,677 & - & - & - \\ \cline{2-8}
4 & RAM (KB) & 13.460 & 1,617.082 & 284,596.043 & - & - & - \\ \cline{2-8}
   & generation time (s)  & 0.570 & 122.966 & 18,728.922 & - & - & - \\ \cline{2-8}
   & search time (s)   & 0.603 & 71.420 & 12,853.887 & - & - & - \\ \hline
\end{tabular}
\end{table}

The disparity in terms of both search time and generation time between different numbers of qubits is likely a function of the increasing complexity of matrix multiplication. In particular, searching requires more matrix multiplications than generation since we generate unitaries for both $S_{i-1}$ and $S_i$ (since only equivalence class representatives are stored, circuits in $S_i$ may not have prefixes in $S_{i-1}$).

\begin{figure}[h]
\centering
\subfloat[Controlled $X$ (depth $1$).]{
	\makebox[0.24\textwidth]{
		\includegraphics[scale=0.35]{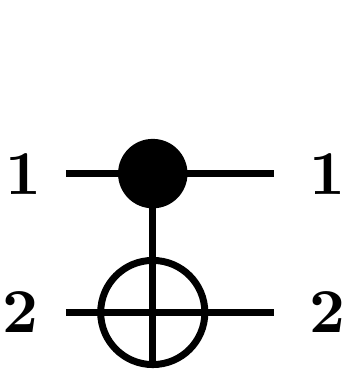}
	}
}
\;
\subfloat[Controlled $Y$ (depth $3$).]{
	\subfloat{
		\includegraphics[scale=0.35]{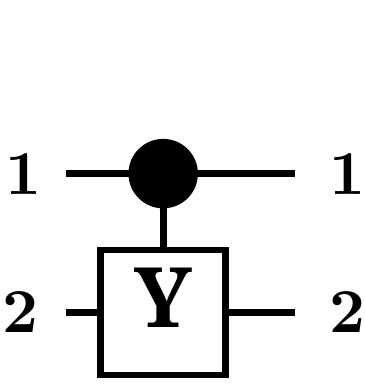}
	}
	\raisebox{2.3ex}{$\equiv$\hspace{-1.4mm}}
	\subfloat{
		\includegraphics[scale=0.35]{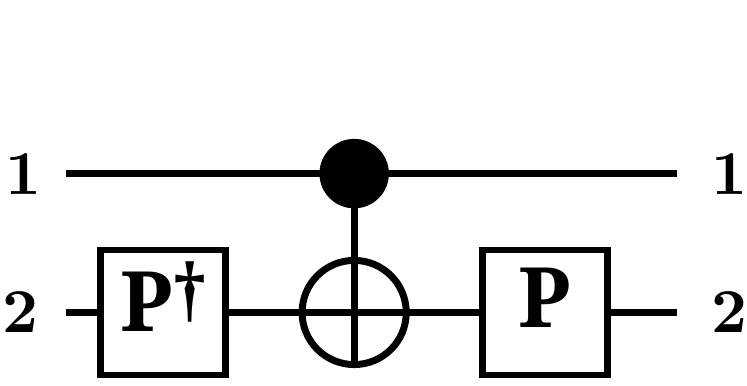}
	}
	\addtocounter{subfigure}{-2}
}
\;
\subfloat[Controlled $Z$ (depth $3$).]{
	\subfloat{
		\raggedleft	
		\includegraphics[scale=0.35]{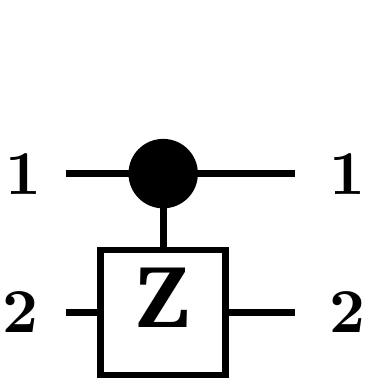}
	}
	\raisebox{2.3ex}{$\equiv$\hspace{-1.4mm}}
	\subfloat{
		\centering
		\includegraphics[scale=0.35]{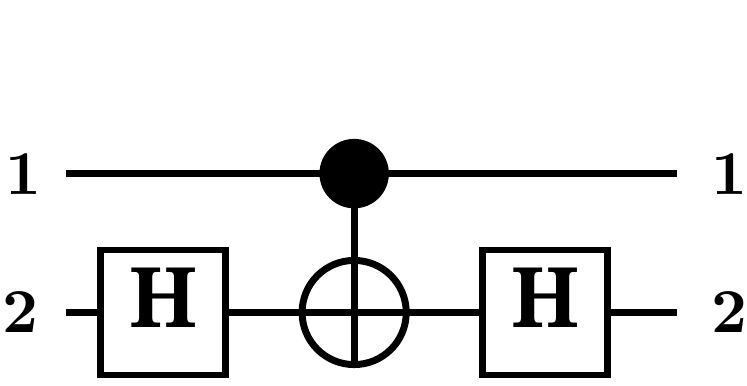}
	}
	\addtocounter{subfigure}{-2}
}
\caption{Controlled Paulis. The $T$-depth of all these circuits is equal to $0$.}
\label{fig:paulis}
\end{figure}

\begin{figure}[h]
\centering
\subfloat[Controlled-$H$ ($T$-depth $2$, total depth $7$).]{
	\subfloat{
		\raggedleft
		\includegraphics[scale=0.35]{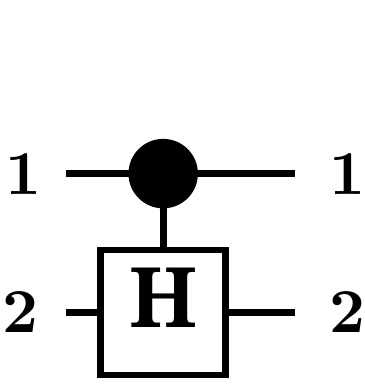}
	}
	\raisebox{2.5ex}{$\equiv$\hspace{-1.4mm}}
	\subfloat{
		\centering
		\includegraphics[scale=0.35]{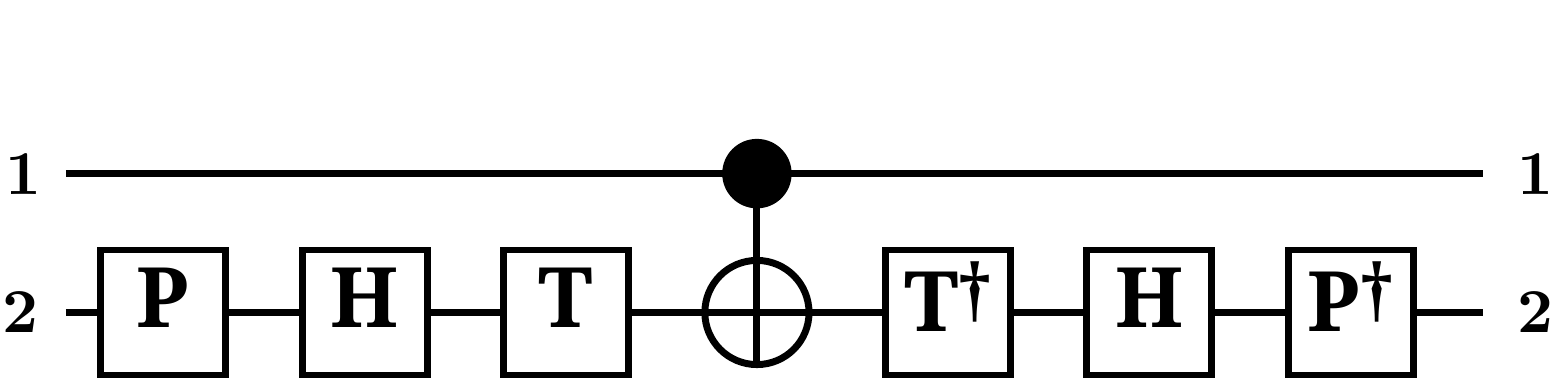}
	}
	\addtocounter{subfigure}{-2}
	\label{fig:CH}
}
\quad
\subfloat[Controlled-$P$ ($T$-depth $2$, total depth $4$).]{
\hspace{2mm}
	\subfloat{
		\raggedleft
		\includegraphics[scale=0.35]{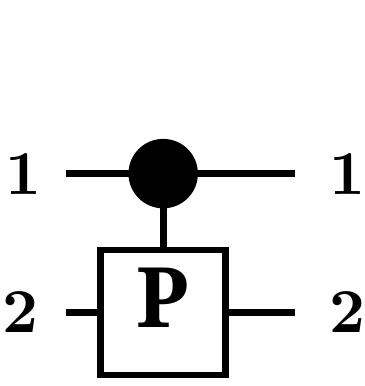}
	}
	\raisebox{2.5ex}{$\equiv$\hspace{-1.4mm}}
	\subfloat{
		\centering
		\includegraphics[scale=0.35]{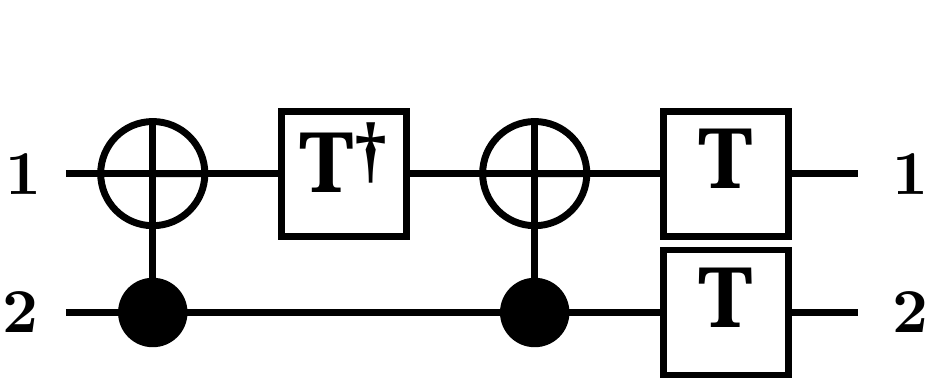}
	}
	\addtocounter{subfigure}{-2}
	\label{fig:CS}
\hspace{2mm}
} \\
\subfloat[Controlled-$\sqrt{X}$ ($T$-depth $2$, total depth $5$).]{
	\hspace{1mm}
	\subfloat{
		\raggedleft
		\includegraphics[scale=0.35]{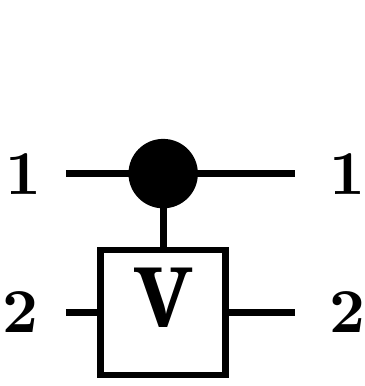}
	}
	\raisebox{2.5ex}{$\equiv$\hspace{-1.4mm}}
	\subfloat{
		\centering
		\includegraphics[scale=0.35]{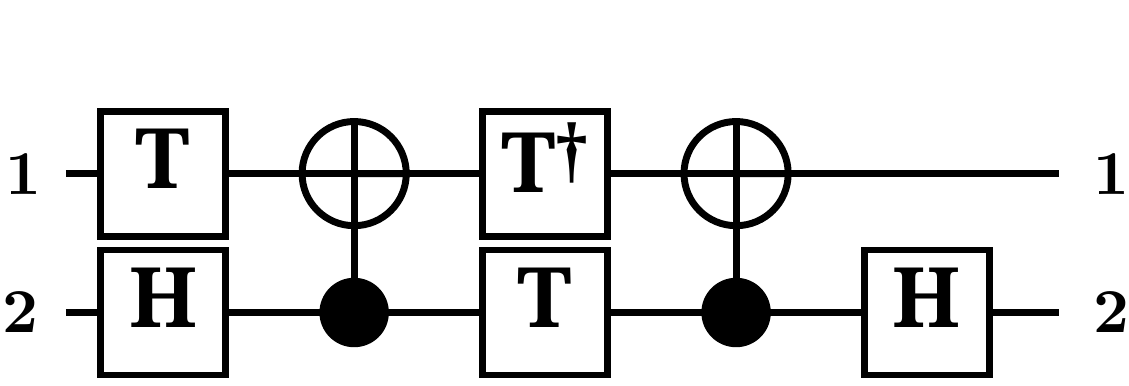}
	}
	\addtocounter{subfigure}{-2}
	\label{fig:CV}
}
\caption{Logical gate implementations of controlled unitaries without ancillas.}
\label{fig:controlledU}
\end{figure}

We performed searches for various 2, 3, and 4 qubit logical operations, using pre-computed databases of circuits. Minimal depth implementations of the singly controlled\footnote{Throughout, a singly controlled-$U$ corresponds to the unitary operator $|0\rangle\langle0|\otimes I + |0\rangle\langle0|\otimes U$.} versions of $H$, $P$, and $V=\sqrt{X}=\frac{1}{2}\begin{pmatrix} 1+i & 1-i \\ 1-i & 1+i \end{pmatrix}$ were computed (Figure~\ref{fig:controlledU}), along with the controlled-$Z$ and $Y$ for completeness (Figure~\ref{fig:paulis}).
\vspace{-8mm}

\begin{figure}[h]
\centering
\subfloat{
	\raggedleft
	\includegraphics[scale=0.35]{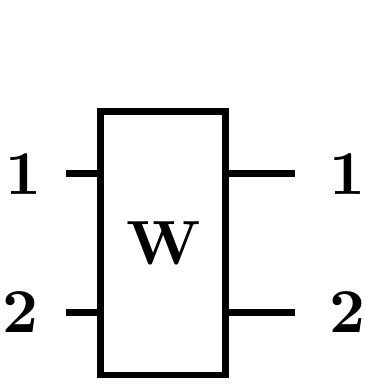}
}
\raisebox{2.5ex}{$\equiv$\hspace{-1.4mm}}
\subfloat{
	\centering
	\includegraphics[scale=0.35]{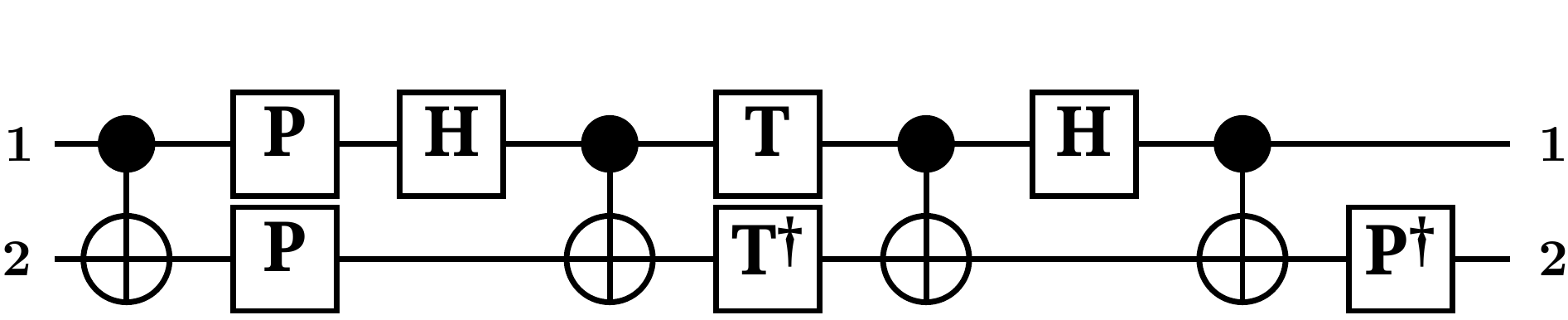}
}
\caption{$W$ gate ($T$-depth $1$, total depth $9$).}
\addtocounter{subfigure}{-2}
\label{fig:W}
\end{figure}

We also optimally decompose (Figure~\ref{fig:W}) the 2-qubit gate $$W=\begin{pmatrix} 1 & 0 & 0 & 0 \\ 0 & \frac{1}{\sqrt{2}} & \frac{1}{\sqrt{2}} & 0 \\ 0 & \frac{1}{\sqrt{2}} & \frac{-1}{\sqrt{2}} & 0 \\ 0 & 0 & 0 & 1 \end{pmatrix}$$ which has found use in at least one interesting quantum algorithm \cite{WELD}.

\begin{figure}[h]
\centering
\subfloat[Toffoli gate ($T$-depth $4$, total depth $8$).]{
	\hspace{-7mm}
	\subfloat{\raisebox{0.2ex}{\includegraphics[scale=0.33]{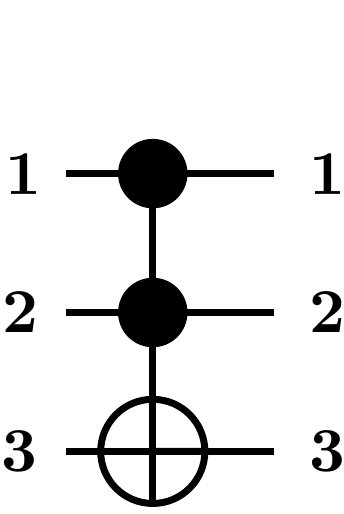}}}
	\raisebox{3.7ex}{\hspace{1.4mm}$\equiv$}
	\subfloat{\includegraphics[scale=0.33]{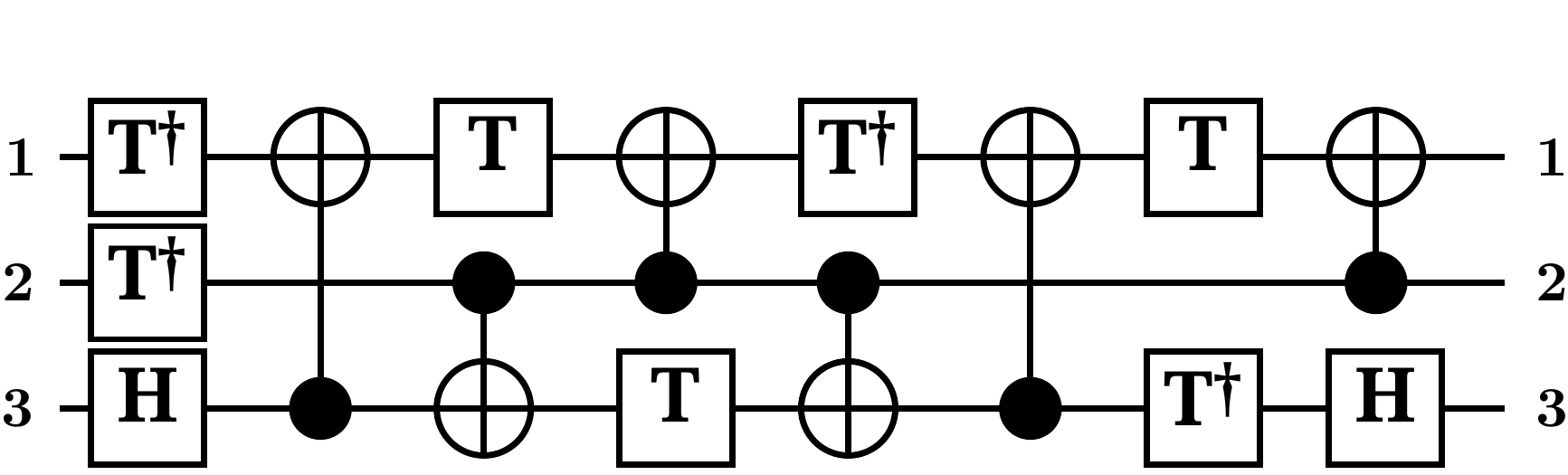}}
	\addtocounter{subfigure}{-2}
	\label{fig:TOF}
}
\subfloat[Toffoli gate, one negative control ($T$-depth $4$, total depth $8$).]{
	\hspace{6mm}
	\subfloat{\raisebox{0.2ex}{\includegraphics[scale=0.33]{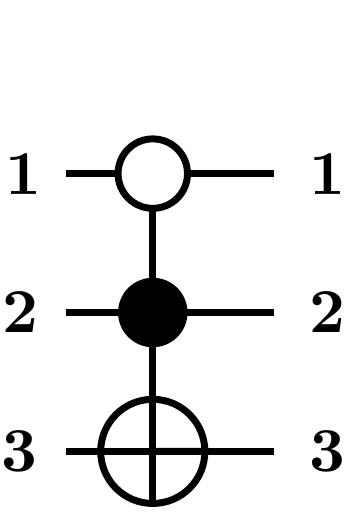}}}
	\raisebox{3.7ex}{\hspace{1.4mm}$\equiv$}
	\subfloat{\includegraphics[scale=0.33]{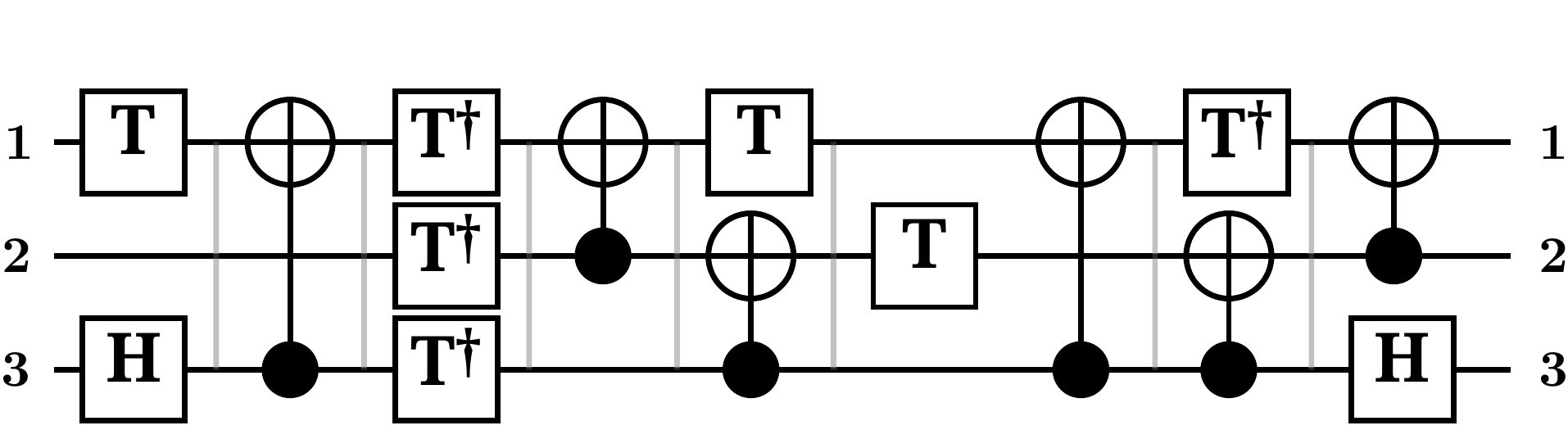}}
	\addtocounter{subfigure}{-2}
	\label{fig:TOF}
	\hspace{6mm}
}

\subfloat[Quantum OR gate ($T$-depth $4$, total depth $8$).]{
	\subfloat{\raisebox{0.15ex}{\includegraphics[scale=0.33]{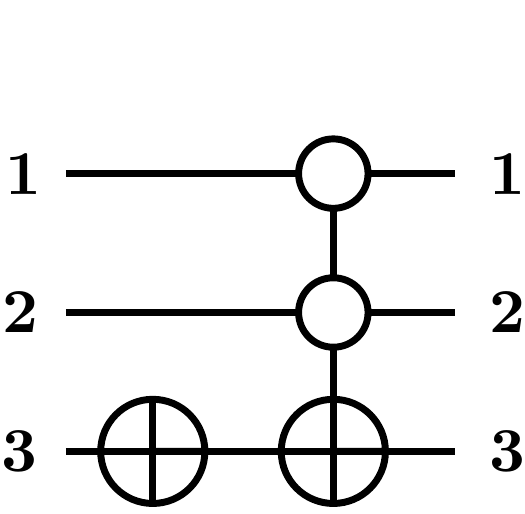}}}
	\raisebox{3.7ex}{\hspace{1.4mm}$\equiv$}
	\subfloat{\includegraphics[scale=0.33]{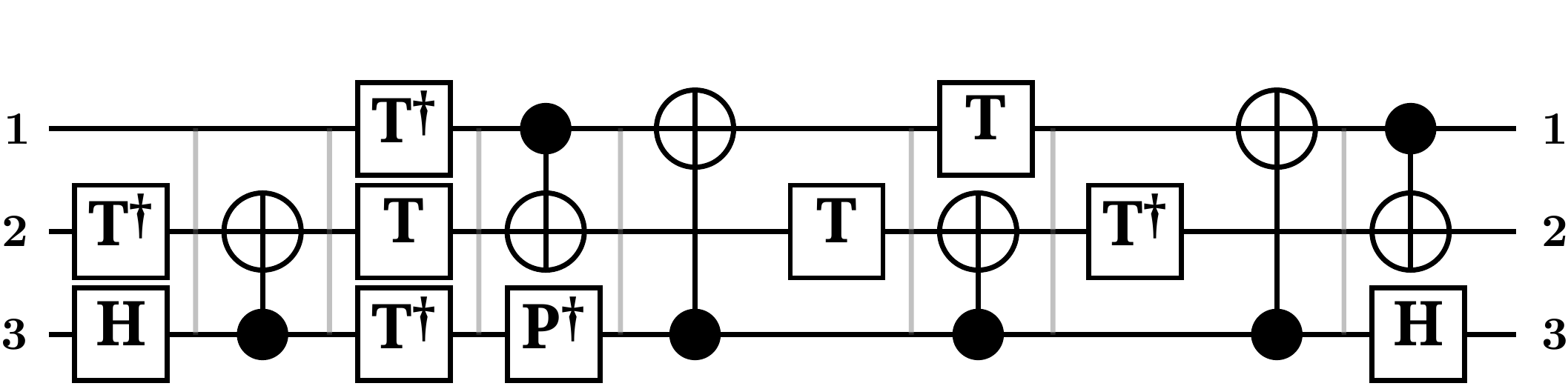}}
	\addtocounter{subfigure}{-2}
	\label{fig:QOR}
}

\subfloat[Peres gate ($T$-depth $4$, total depth $8$).]{
	\subfloat{\raisebox{0.2ex}{\includegraphics[scale=0.33]{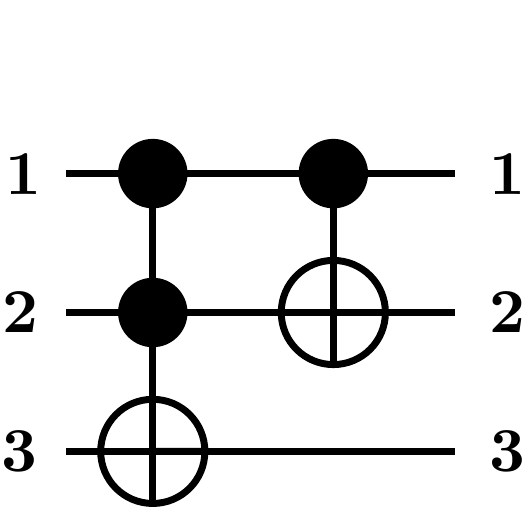}}}
	\raisebox{3.7ex}{\hspace{1.4mm}$\equiv$}
	\subfloat{\includegraphics[scale=0.33]{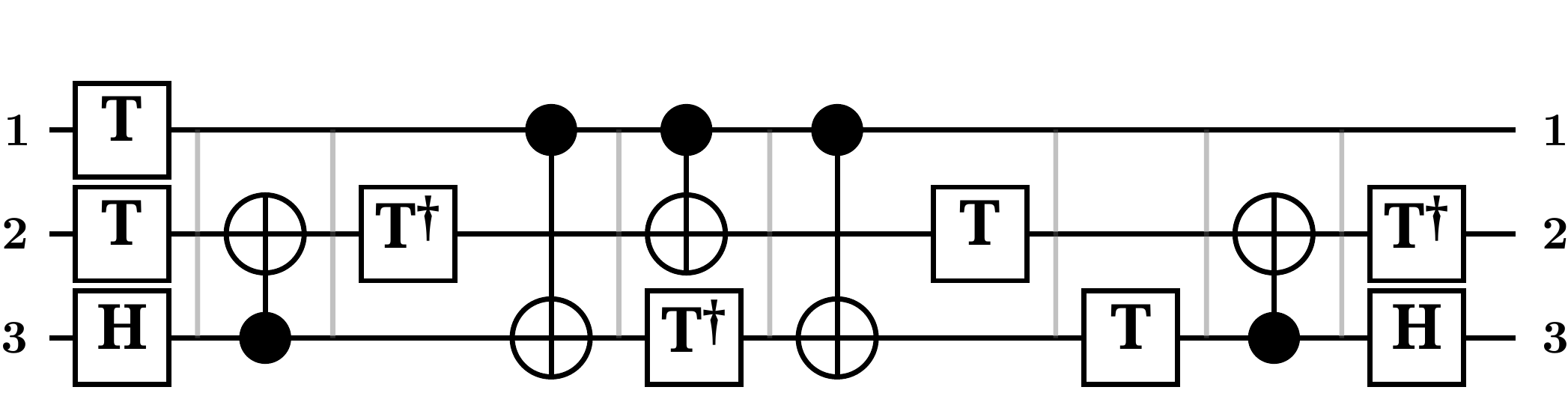}}
	\addtocounter{subfigure}{-2}
	\label{fig:PER}
}

\subfloat[Fredkin gate ($T$-depth $4$, total depth 10).]{
	\subfloat{\raisebox{0.4ex}{\includegraphics[scale=0.33]{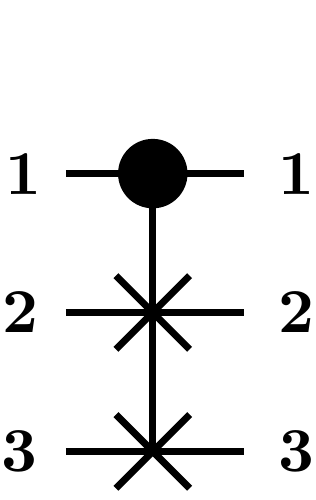}}}
	\raisebox{3.5ex}{\hspace{1.4mm}$\equiv$}
	\subfloat{\includegraphics[scale=0.33]{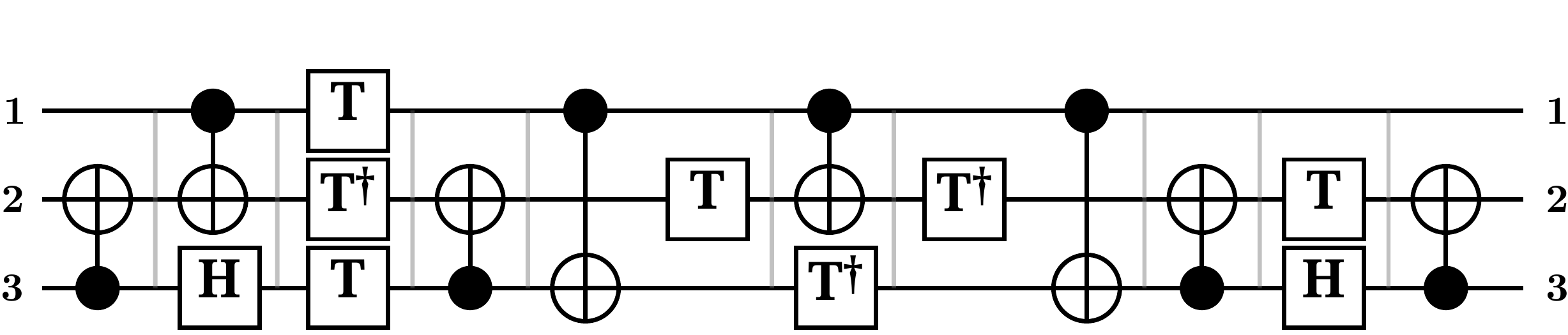}}
	\addtocounter{subfigure}{-2}
	\label{fig:FRED}
}
\caption{3-qubit logical gates with no ancillas.}
\label{fig:threequbits}
\end{figure}

Some 3-qubit unitaries with minimal depth implementations found include the well-known Toffoli gate (controlled-$CNOT$), Fredkin gate (controlled-$SWAP$), quantum OR (defined as the unitary mapping $|a\rangle|b\rangle|c\rangle\mapsto|a\rangle|b\rangle|c\oplus a\vee b\rangle$), and Peres gate \cite{PER} (Figure~\ref{fig:threequbits}). It should be noted our circuit reduces the total depth of the Toffoli gate from 12 \cite{SK} to 8.

\begin{figure}[h]
\centering
\subfloat[Controlled-$P$ ($T$-depth $1$, total depth $5$).]{
	\makebox[0.5\textwidth]{
	\subfloat{
		\raisebox{0.8ex}{\includegraphics[scale=0.33]{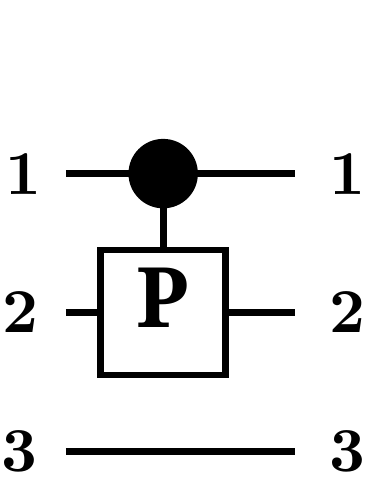}}
	}
	\raisebox{3.7ex}{$\equiv$\hspace{-1.4mm}}
	\subfloat{
		\includegraphics[scale=0.33]{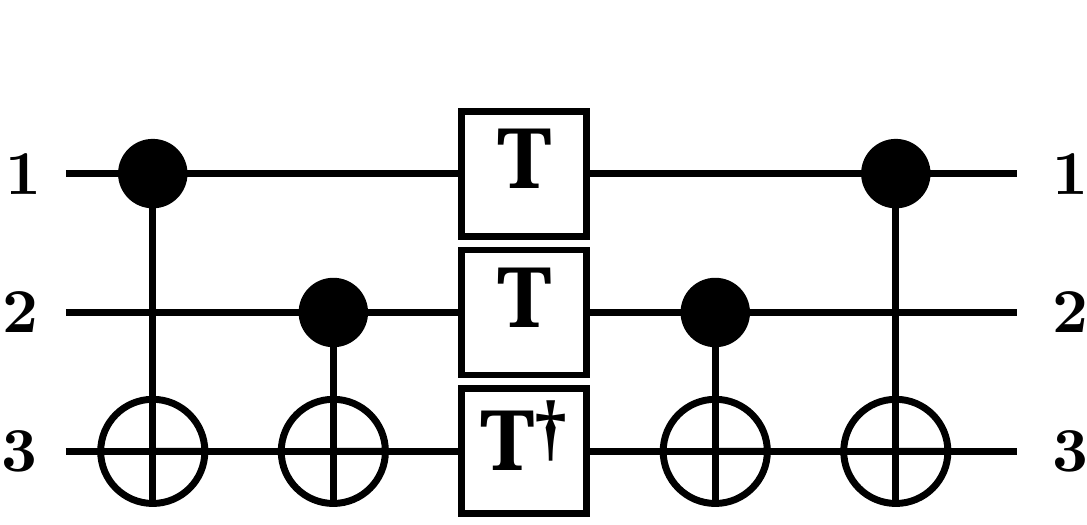}
	}
	\addtocounter{subfigure}{-2}
	\label{fig:CSa}
	}
}
\subfloat[Controlled-$\sqrt{X}$ ($T$-depth $1$, total depth $5$).]{
	\subfloat{
		\raisebox{0.8ex}{\includegraphics[scale=0.33]{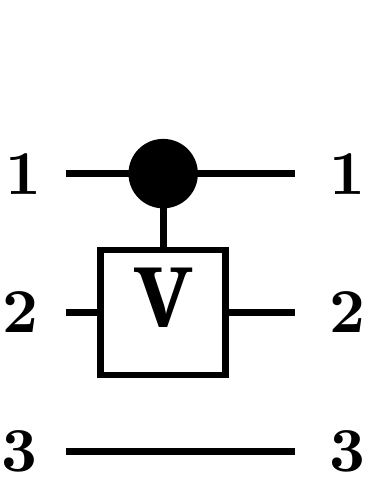}}
	}
	\raisebox{3.7ex}{$\equiv$\hspace{-1.4mm}}
	\subfloat{
		\includegraphics[scale=0.33]{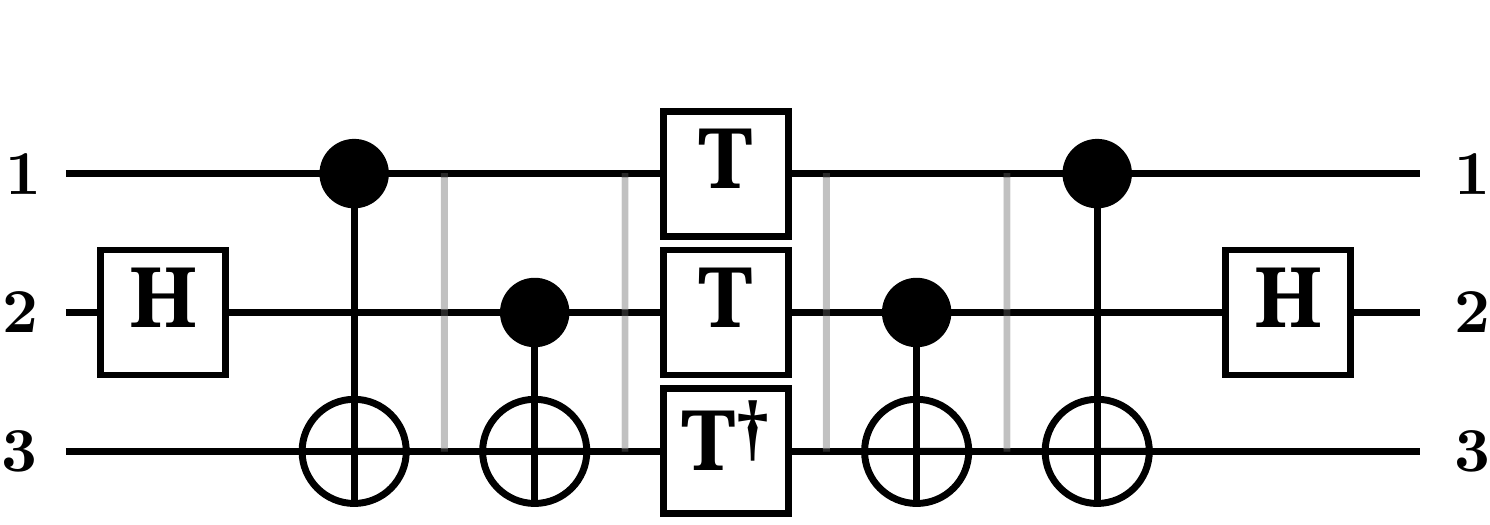}
	}
	\addtocounter{subfigure}{-2}
	\label{fig:CVa}
}
\caption{Reduced $T$-depth implementations utilizing ancillas.  Note that qubit 3 is initialized in and returned to state $|0\rangle$.}
\label{fig:CSancilla}
\end{figure}

\begin{figure}[h]
\centering
\subfloat{\raisebox{0.8ex}{\includegraphics[scale=0.32]{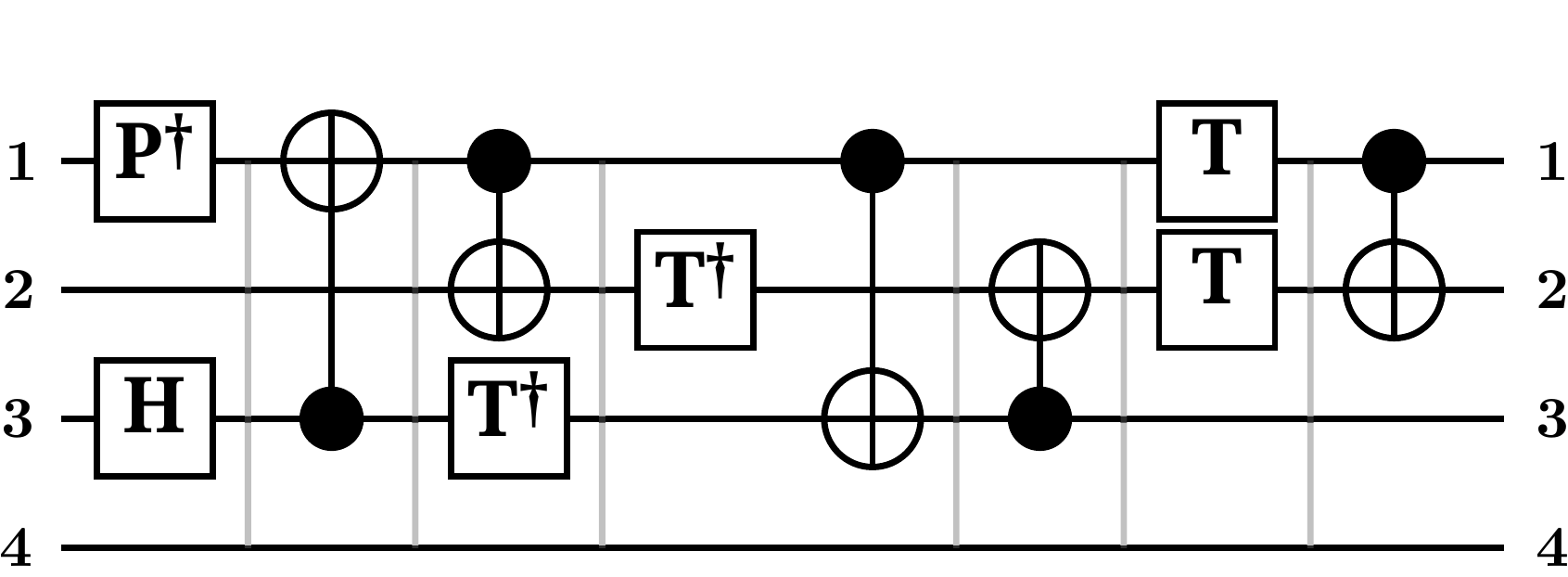}}}
\raisebox{4.9ex}{\hspace{1.4mm}$\equiv$}
\subfloat{\includegraphics[scale=0.32]{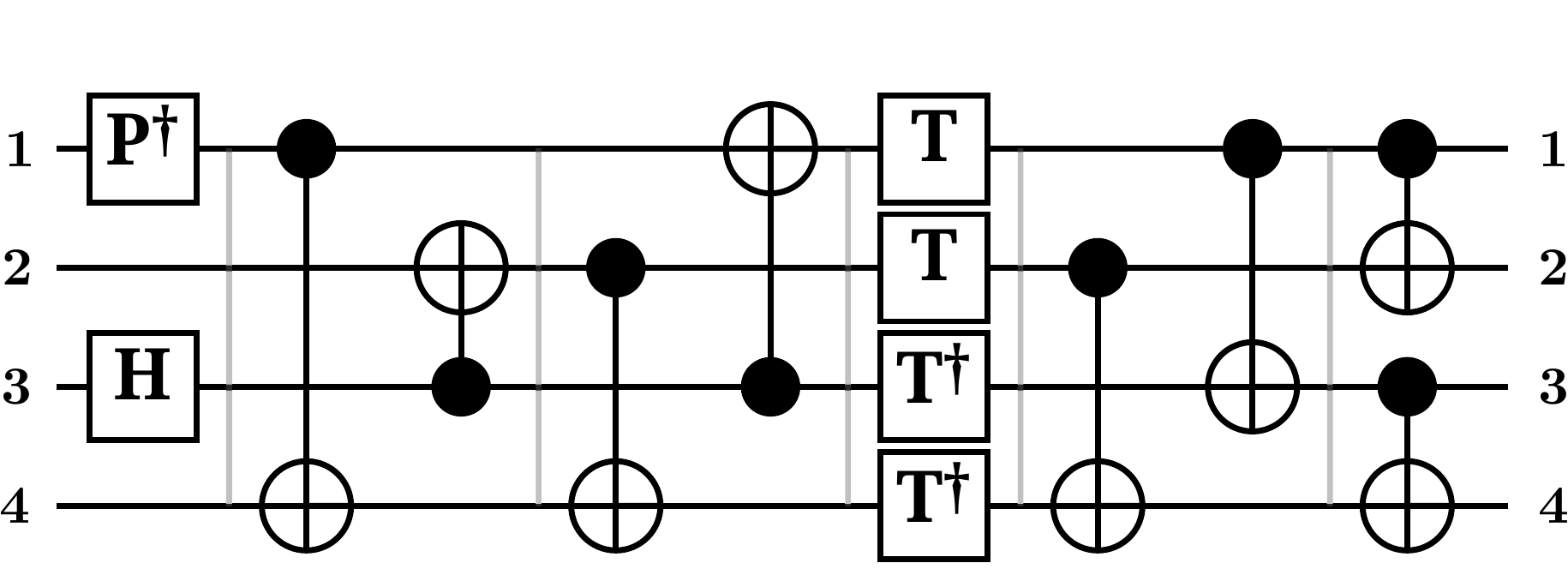}}
\caption{Addition of one ancilla (qubit 4), initialized and returned in state $|0\rangle$, reduces the minimum circuit depth from 7 (left) to 6 (right).}
\label{fig:reduceddepth}
\end{figure}

Searches were also performed for each of the above $n$-qubit gates using up to $4-n$ ancillas -- these searches were performed up to the maximum depth for the total number of qubits, as seen in Table 1. None of the logical gates tested were found to admit circuits with shorter depth or fewer $T$ gates, though circuits for controlled-$P$ and controlled-$\sqrt{X}$ were found with smaller $T$-depth (Figure~\ref{fig:CSancilla}). Additionally, a circuit was found that did have reduced minimal depth when decomposed using an ancilla (Figure~\ref{fig:reduceddepth}), together with the reduced $T$-depth circuits providing clear motivation for the use of ancillas to optimize circuit execution time.

Among other gates attempted were the 3-qubit quantum Fourier transform, which was proven to have no circuit in our instruction set with depth at most 10, and the 4-qubit Toffoli gate (controlled-Toffoli) and 1-bit full adder, with no circuits of depth at most 6. Additionally, both the controlled-$T$ and controlled-$\sqrt[4]{X}$ gates were proven to have no implementations of depths at most 10 or 6 using one or two ancillas, respectively.

\begin{figure}[h]
\centering
\subfloat{\raisebox{0.8ex}{\includegraphics[scale=0.33]{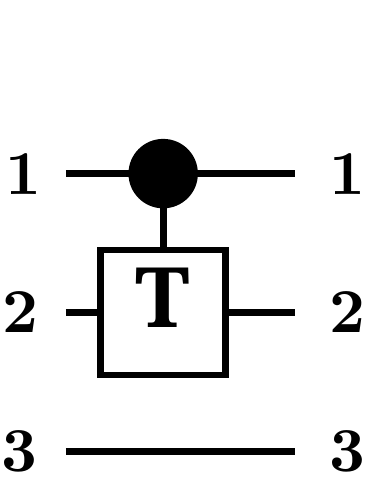}}}
\raisebox{3.6ex}{\hspace{1.4mm}$\equiv$}
\subfloat{\includegraphics[scale=0.33]{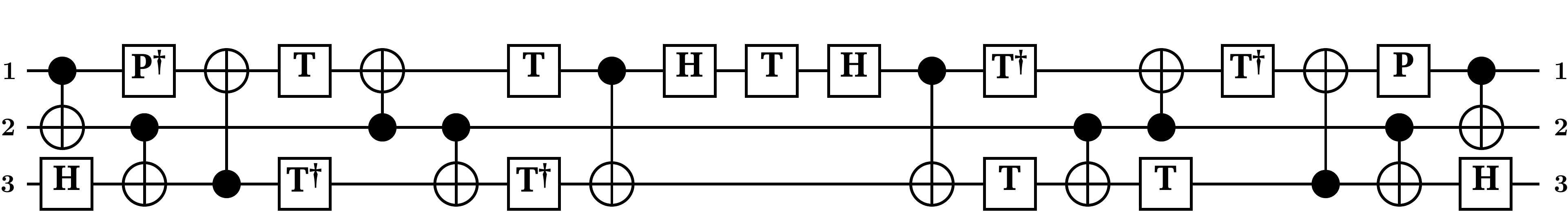}}
\caption{Circuit implementing a controlled-$T$ gate ($T$-depth $5$, total depth $19$).  Note that qubit 3 is initialized in and returned to state $|0\rangle$.}
\label{fig:controlledT}
\end{figure}

\begin{figure}[h]
\centering
\subfloat{\raisebox{0.2ex}{\includegraphics[scale=0.32]{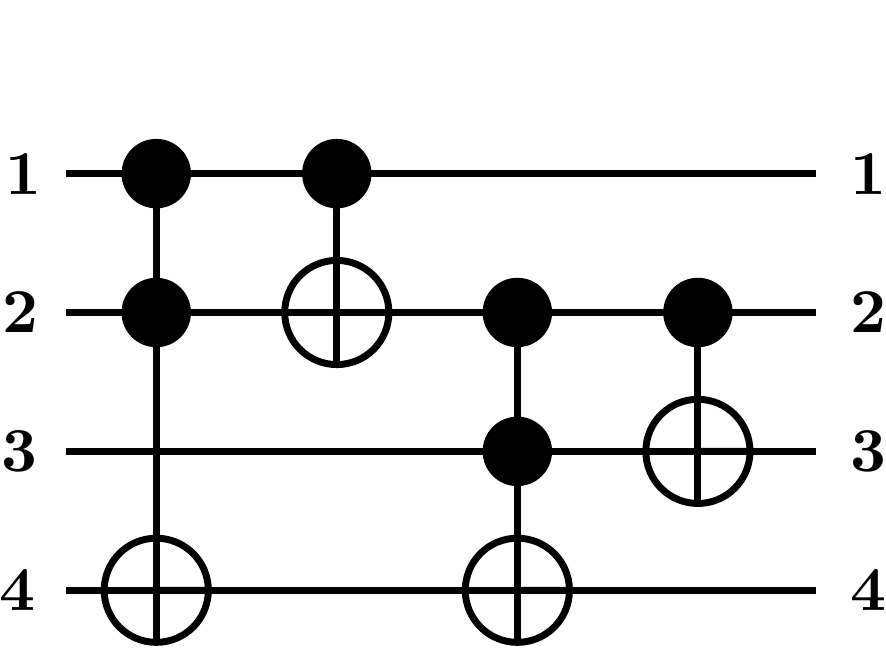}}}
\raisebox{4.9ex}{\hspace{1.4mm}$\equiv$}
\subfloat{\includegraphics[scale=0.32]{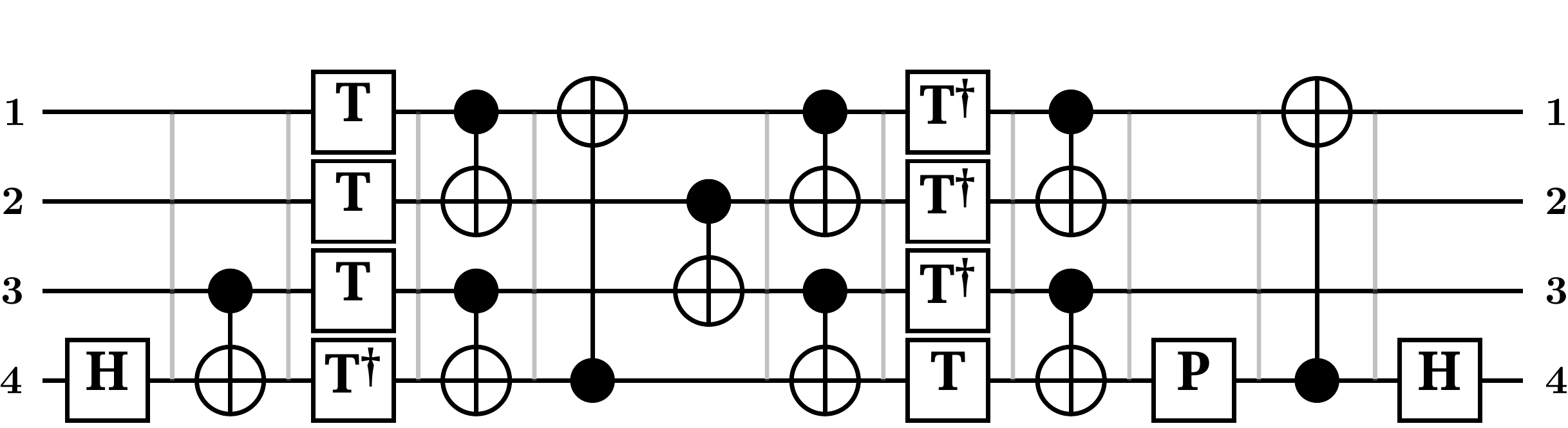}}
\caption{Circuit implementing a reversible 1-bit full adder.}
\label{fig:adder}
\end{figure}

We did however optimize a known circuit implementing the controlled-$T$ gate, as well as one implementing a 1-bit full adder using our algorithm. Specifically, we generated a circuit for the controlled-$T$ gate using the decomposition $FREDKIN\cdot (I\otimes I\otimes T) \cdot FREDKIN$, and a circuit for the 1-bit adder by using the implementation found in \cite{FEYN}, substituting the circuit in Figure~\ref{fig:PER} for the Peres gate. Then we performed a peep-hole optimization by taking small subcircuits and replacing them with shorter, lower gate count circuits synthesized using our algorithm. The circuit for controlled-$T$, shown in Figure~\ref{fig:controlledT}, reduces the number of $T$ gates from 15 to 9, $CNOT$ gates from 16 to 12, and $T$-depth from 9 to 5, while the 1-bit adder circuit (Figure~\ref{fig:adder}) reduces the number of $T$ gates from 14 to 8, $CNOT$ gates from 12 to 10, $H$ gates from 4 to 2, and $T$-depth from 8 to 2. These results provide strong evidence for the effectiveness of peep-hole re-synthesis as a full-scale optimization tool.

\subsection{Optimal $T$-depth implementations}

Experiments were also performed to find circuits with minimum $T$-depth, using the modified algorithm. The bottleneck in this case is the sheer size of the Clifford group, both troublesome for generating the group itself, and for increasing the $T$-depth when searching. For 2 qubits, generation of the 11,520 unique Clifford group elements (up to phase) required approximately 1 second of computing time and less than 2 seconds to search for a unitary up to 1 $T$-stage, or 2 $T$-stages and ending in a non-Clifford operation. In practice, this was enough to find minimum $T$-depth implementations of the 2-qubit gates in question. By contrast, generation of the 92,897,280 unique 3-qubit Clifford group elements required almost 4 days to compute.

\begin{figure}[h]
\centering
\subfloat{
	\raggedleft
	\includegraphics[scale=0.35]{figures/CHpic-eps-converted-to}
}
	\raisebox{2.3ex}{$\equiv$\hspace{-1.4mm}}
\subfloat{
	\includegraphics[scale=0.35]{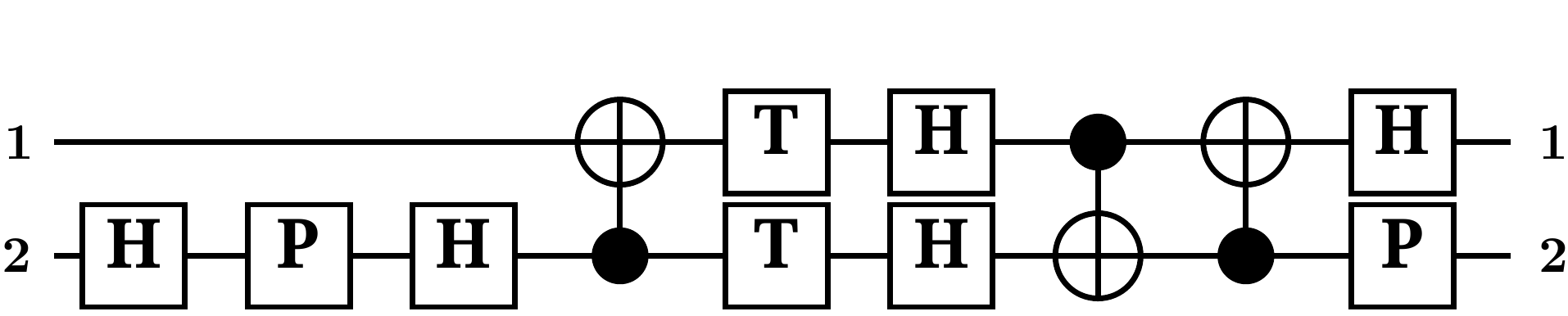}
}
\caption{Circuit implementing a controlled-$H$ gate ($T$-depth $1$, total depth $9$).}
\label{fig:CH-T}
\end{figure}

The minimal $T$-depth controlled-$H$ gate (Figure~\ref{fig:CH-T}) required less than one second to compute, after generating the Clifford group. Minimal $T$-depth circuits for other 2-qubit logical gates were not found to decrease the number of $T$-stages compared to the minimal depth circuits, and thus the circuits shown for controlled-$P$, controlled-$\sqrt{X}$, and $W$ are optimal both in circuit depth and $T$-depth. As a result, allowing the use of ancillas can strictly decrease the minimum $T$-depth required to implement a given unitary, since implementations of the controlled-$P$ and $\sqrt{X}$ gates with ancillas were found with lower $T$-depth.

\begin{figure}[h]
\centering
\subfloat{
	\includegraphics[scale=0.33]{figures/TOFpic-eps-converted-to}
}
	\raisebox{3.5ex}{$\equiv$\hspace{-1.4mm}}
\subfloat{
	\includegraphics[scale=0.33]{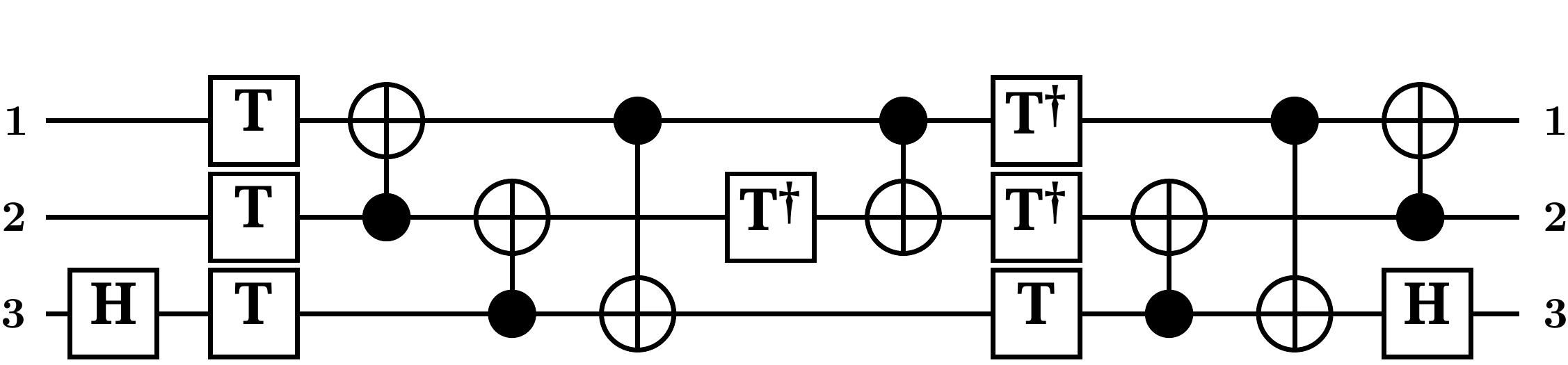}
}
\caption{Circuit implementing a Toffoli gate ($T$-depth 3, total depth 9).}
\label{fig:depth3toff}
\end{figure}

While no Toffoli has yet been found with provably minimal $T$-depth and zero ancilla, a circuit with $T$-depth 3 implementing the Toffoli gate (Figure~\ref{fig:depth3toff}) has been found using our main algorithm; as the Toffoli appears to require a minimum of $7$ $T$ gates to implement, we conjecture this is minimal. Furthermore, it reduces the number of $T$-stages from 5 \cite{SK} to 3, providing an approximate 40\% speed-up in fault tolerant architectures where Clifford group gates have negligible cost compared to the $T$ gate.

While these circuits are maximally parallelized with $T$-depth $\lceil m/n \rceil$ for $m$ $T$-gates and $n$ qubits, in general not every circuit can be compressed in such a way. 

\subsection{Exact decomposition of controlled unitaries}

It is a well-known fact that the controlled version of any circuit can be generated by replacing each individual gate with a controlled version of that gate, with the control qubit of the entire circuit functioning as the control qubit of each gate \cite{KLM}. The minimal-depth circuits computed in Section \ref{sec:performance} allow us 
to establish the following result. 

\begin{theorem}
Let the gate cost of a circuit be given by a vector $\mathbf{x}=[x_H, x_P, x_{C}, 
x_T]^t$, where $x_H$ denotes the number of $H$ gates, $x_P$ denotes the number of $P$-gates or $P^\dagger$-gates, $x_{C}$ denotes the number of $CNOT$ gates, and $x_T$ denotes the number of $T$-gates or $T^\dagger$-gates. Suppose $U$ can be implemented to error $\epsilon\geq 0$ by a circuit over $\mathcal{G}=\{ H, P, P^\dagger, CNOT, T, T^\dagger\}$ with gate cost of $\mathbf{x}$. Then controlled-$U$ can be implemented to error at most $\epsilon$ over $\mathcal{G}$ by a circuit of gate cost $A\mathbf{x}$, where $$A=\begin{bmatrix} 2 & 0 & 2 & 4 \\ 2 & 0 & 0 & 2 \\ 1 & 2 & 6 & 12 \\ 2 & 3 & 7 & 9 \end{bmatrix}.$$ The circuit for controlled-$U$ uses exactly one ancilla qubit if one or more $T$-gates are present in the decomposition of $U$, and no ancilla qubits otherwise. Furthermore, controlled-$U$ can be implemented in a $T$-depth of at most $x_H + 2 x_P + 3 x_{C} + 5 x_T$.
\end{theorem}
\noindent
{\em Proof:} 
Assume that $U$ admits an $\epsilon$-approximation over $\mathcal{G}$ with associated cost vector $\mathbf{x}=[x_H, x_P, x_{C}, x_T]^t$. As shown in Section \ref{sec:performance}, for each gate $H$, $P$, $CNOT$, $T$, the corresponding singly controlled gate can be implemented exactly over $\mathcal{G}$. Specifically, we obtain from Figure \ref{fig:controlledU} for each controlled-$H$ gate a cost of $[2,2,1,2]$ and for each controlled-$P$ gate a cost of $[0,0,2,3]$. From Figures \ref{fig:threequbits} and \ref{fig:controlledT} we obtain costs for each controlled-$CNOT$ (i.e., Toffoli) and each controlled-$T$ gate of $[2,0,6,7]$ and $[4,2,12,9]$, respectively. Since the total cost is linear in the costs of the gates, we obtain that claimed total cost of $A \mathbf{x}$. The approximation error $\epsilon$ is unchanged as compared to $U$ since no further errors are introduced in the factorization. Finally, the claimed bound for the overall $T$-depth holds since the $T$-depths of each $H$, $P$, $CNOT$, and $T$ gates can be upper bounded by $1$, $2$, $3$, and $5$, respectively, where here we used the $T$-depth $3$ circuit in Figure \ref{fig:depth3toff} to derive an upper bound for the Toffoli gate.
\hfill $\Box$

\subsection{$T$ gate parallelization}

It was noted earlier that the three $T$ gates used in the controlled-$P$ and controlled-$\sqrt{X}$ circuits (Figures~\ref{fig:CS},~\ref{fig:CV}) can be parallelized to $T$-depth 1 using one ancilla (Figure~\ref{fig:CSancilla}). Similarly, the seven $T$ gates in the Toffoli decomposition (Figure~\ref{fig:depth3toff}) can be parallelized to $T$-depth 1 using 4 ancilla \cite{SEL}. The parallelized $T$ gates were separated by networks of $CNOT$ gates in each of these cases. We prove a theorem relating the number of $T$ gates in a $\{CNOT, T\}$ circuit, and the achievable $T$-depth for a given number of ancilla.

\begin{figure}[h]
\centering
\subfloat{\raisebox{0.8ex}{\includegraphics[scale=0.32]{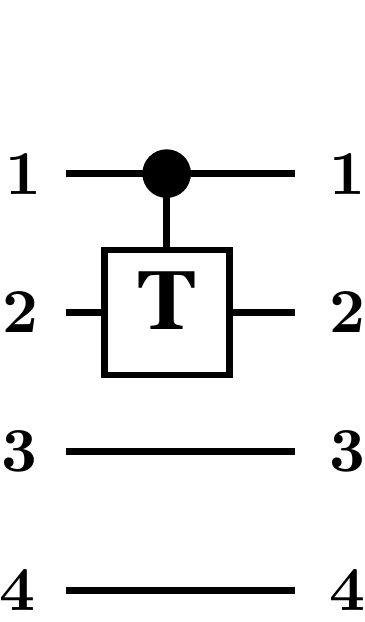}}}
\raisebox{5.0ex}{\hspace{1.4mm}$\equiv$}
\subfloat{\includegraphics[scale=0.32]{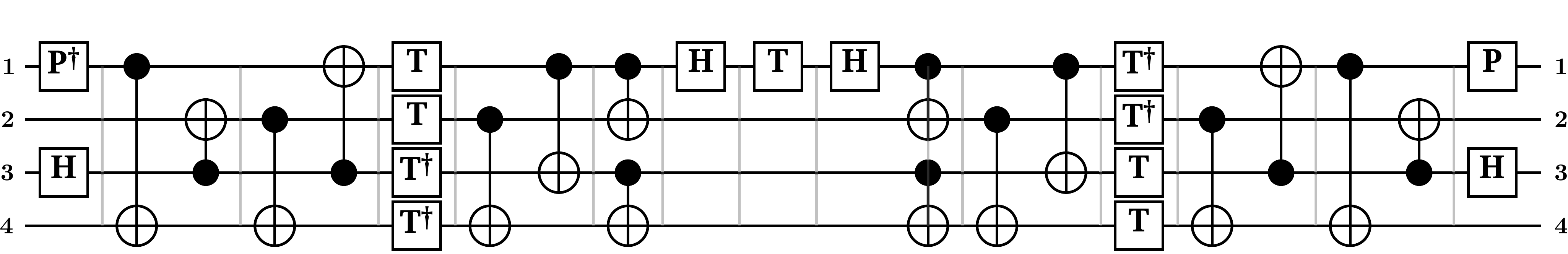}}
\caption{$T$-depth 3 implementation of the controlled-$T$ gate.}
\label{fig:controlledTancilla}
\end{figure}

\begin{theorem}
\label{thm:Tpar}
Any circuit on $n$ qubits over $\{CNOT, T\}$ with $k$ $T$ gates can be implemented by a circuit over $\{CNOT, T\}$ on $n$ qubits and $m$ ancilla, initialized and returned in state $|0\rangle$, with $T$-depth at most $\left\lceil \frac{k}{m+1} \right\rceil$.
\end{theorem}

Before proving Theorem~\ref{thm:Tpar}, we first prove a simple lemma.
\begin{lemma}
\label{lem:Tpar}
Unitary $U\in U(2^n)$ is exactly implementable by an $n$-qubit circuit over $\{CNOT, T\}$ with $k$ $T$ gates if and only if $$U|a_1a_2...a_n\rangle = \omega^t|g(a_1, a_2, ..., a_n)\rangle$$ where $\omega=e^{\frac{i\pi}{4}}$ and $t=f_1(a_1, ..., a_n) + f_2(a_1, ..., a_n) + \cdots + f_k(a_1, ..., a_n)$ for some linear Boolean functions $f_1, f_2, ..., f_k$ and linear reversible function $g$.
\end{lemma}

\noindent
{\em Proof:} We remind readers that in the computational basis, $CNOT:|a\rangle|b\rangle\mapsto |a\rangle|b\oplus a\rangle$ and $T:|a\rangle\mapsto \omega^a|a\rangle$.

The forward direction can be observed by writing the circuit implementing $U$ as an alternating product of $CNOT$ and $T$ circuits. Each $CNOT$ circuit computes a linear reversible function $f$ on the inputs, while a following $T$ gate on the $i$th qubit adds an overall phase multiple of $\omega^{f^i(a_1, ..., a_n)}$ where $f^i$ denotes the linear Boolean function corresponding to the $i$th output of $f$. Since the overall phase has no effect on linear reversible functions, the functions computed by each $CNOT$ circuit compose and can be written as functions of the inputs, completing the proof.

The reverse direction is equally simple by noting that for any linear Boolean function $f_i$ on $n$ inputs, $f_i$ is an output of some linear reversible function on $n$ inputs, and thus can be computed using only $CNOT$ gates \cite{P1}. By applying a $T$ gate to the qubit with state $|f_i(a_1,a_2,...,a_n)\rangle$ then uncomputing, the input state is recovered, with an added phase multiple of $\omega^{f_i(a_1, a_2, ..., a_n)}$. It then suffices to observe that $g$ can be computed with $CNOT$ gates, so that $U$ can be implemented with $CNOT$s and $k$ $T$ gates.
\hfill $\Box$

We now proceed to prove Theorem~\ref{thm:Tpar}.

\noindent
{\em Proof of Theorem~\ref{thm:Tpar}:} 
Suppose $U$ is implementable by a circuit over $\{CNOT, T\}$ using $k$ $T$ gates. Then $$U|a_1a_2...a_n\rangle = \omega^t|g(a_1, a_2, ..., a_n)\rangle$$ where $t=f_1(a_1, ..., a_n) + f_2(a_1, ..., a_n) + \cdots + f_k(a_1, ..., a_n)$ for some linear Boolean functions $f_1, f_2, ..., f_k$ and linear reversible function $g$. 

Consider $k\leq m$ and let $f$ be defined as $$f|a_1\cdots a_n\rangle|b_1\cdots b_m\rangle=|a_1\cdots a_n\rangle|c_1\cdots c_k\rangle|b_{k+1}\cdots b_m\rangle,$$ where $c_i=b_i\oplus f_i(a_1, a_2, ..., a_n)$. Indeed, $f$ is linear since each $f_i$ is a linear Boolean function and reversible since $f=f^{-1}$, so $f$ is computable by some quantum circuit over $\{CNOT\}$. We then easily observe that if $V=I^{\otimes n}\otimes T^{\otimes k}\otimes I^{\otimes m-k},$ $$f^{-1}Vf|a_1a_2\cdots a_n\rangle|0\rangle^{\otimes m} =\omega^t|a_1a_2\cdots a_n\rangle|0\rangle^{\otimes m}.$$ Since $g$ is a linear reversible function, $U$ can thus be implemented by a circuit over $\{CNOT, T\}$ in $T$-depth $1=\left\lceil \frac{k}{m+1} \right\rceil$.

Now suppose $k > m$. As before, there exists a linear reversible function $f$ implemented by a circuit over $\{CNOT\}$ such that $$f|a_1a_2\cdots a_n\rangle|b_1b_2\cdots b_m\rangle=|a_1a_2\cdots a_n\rangle|c_1c_2\cdots c_m\rangle,$$ where $c_i=b_i\oplus f_i(a_1, a_2, ..., a_n)$. Additionally, $f_{m+1}$ is an output of some linear reversible function $h$, so the first $m+1$ factors of $\omega$ can be computed in $T$-depth 1 by implementing the unitary $f^{-1}h^{-1}Vhf,$ where $V$ is a tensor product of $I$ and $m+1~T$ gates.

As a result, $(U\otimes I^{\otimes m})|a_1a_2\cdots a_n\rangle|0\rangle^{\otimes n}=(U'\otimes I^{\otimes m})f^{-1}h^{-1}Vhf|a_1a_2\cdots a_n\rangle|0\rangle^{\otimes n},$ where $U'|a_1a_2...a_n\rangle = \omega^{t'}|g(a_1, a_2, ..., a_n)\rangle$ and $t'=f_{m+2}(a_1, ..., a_n) + \cdots + f_k(a_1, ..., a_n)$. By Lemma~\ref{lem:Tpar} $U'$ can be implemented by a circuit over $\{CNOT, T\}$ with $k-(m+1)$ $T$ gates, and thus $U$ can be implemented in $T$-depth at most $\left\lceil \frac{k}{m+1} \right\rceil$ by induction.
\hfill $\Box$

In general we can do much better than $T$-depth $\left\lceil \frac{k}{m+1} \right\rceil$, as on average more than one $f_i$ can be computed reversibly into data qubits at a time. Specifically, whenever there are $l$ linearly independent functions to be computed, $l$ data qubits can be used. The possible $T$-depth given $n$ data qubits and $m$ ancilla is then given by the size of a minimal partition $\{S_1, S_2, ..., S_l\}$ of $\{f_1, f_2, ..., f_k\}$, such that each $|S_i|\leq m+\operatorname{dim}(\operatorname{span}S_i).$

\begin{figure}[h]
\centering
\subfloat{\raisebox{0.8ex}{\includegraphics[scale=0.32]{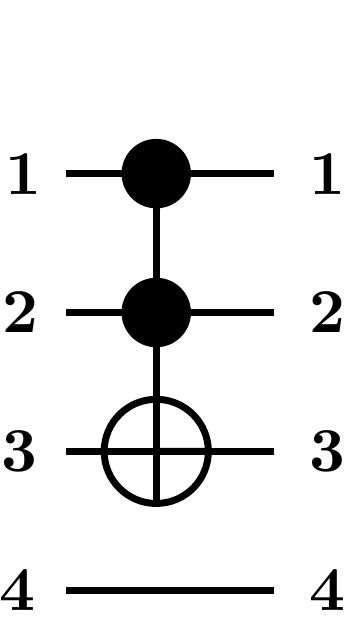}}}
\raisebox{5.0ex}{\hspace{1.4mm}$\equiv$}
\subfloat{\includegraphics[scale=0.32]{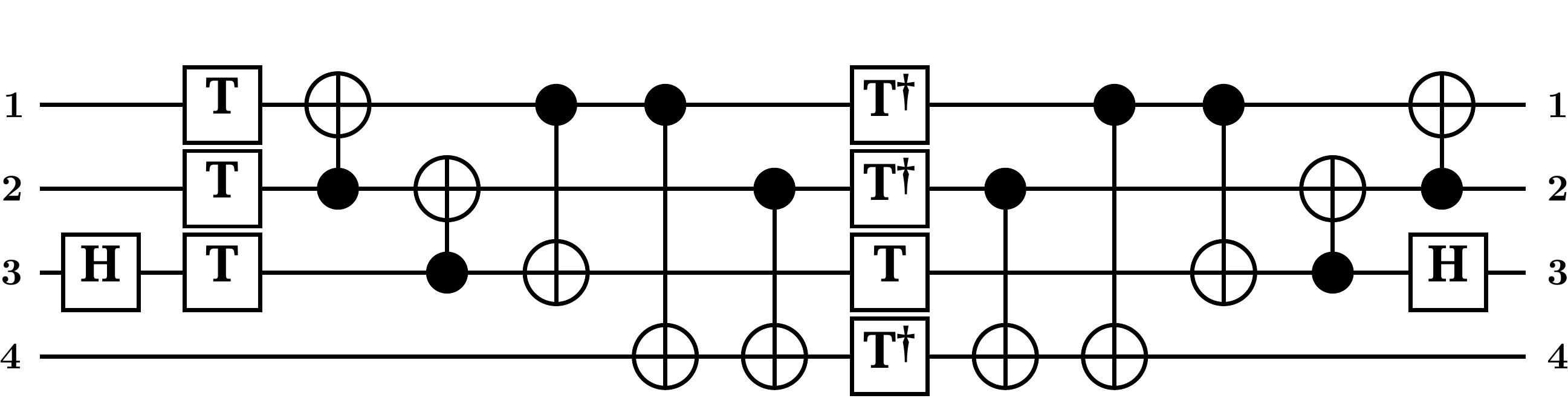}}
\caption{$T$-depth 2 implementation of the Toffoli gate.}
\label{fig:tdepth2toff}
\end{figure}

As an example of $T$ parallelization, we rewrite the $T$-depth 3 Toffoli (Figure~\ref{fig:depth3toff}) in $T$-depth 2 by using one ancilla (Figure~\ref{fig:tdepth2toff}), as well as the implementation of the controlled-$T$ gate (Figure~\ref{fig:controlledT}) in $T$-depth 3 using one ancilla (Figure~\ref{fig:controlledTancilla}). Though we do not give a circuit, the 1-bit full adder circuit in Figure~\ref{fig:adder} can be rewritten in $T$-depth 1 using 4 ancilla.

\section{Conclusion}

In this paper, we have described a simple algorithm for finding a minimal depth quantum circuit implementing a given unitary $U$ in a specific gate set. Our primary focus was to find unitary factorizations with either minimal depth, or a minimal number of sequential non-Clifford group gates. Our computations have found minimal depth circuits for many important logical gates, in some cases providing significant speed-up over known or algorithmically generated circuits. Our implementation takes approximately 32 minutes to generate all equivalence class representatives of 3-qubit circuits up to depth 4 in the instruction set $\{H, P, P^\dagger, CNOT, T, T^\dagger\}$, and 415 seconds to find any optimal 3-qubit circuit up to depth 8.

While these numbers are not at the same level of those found in \cite{M1}, they cannot be directly compared. Golubitsky and Maslov considered 4 bit Boolean functions, which admit a representation via 64 bits that also allows common operations such as permutation and inversion to be carried out by bitwise operations. While we use similar bit twiddling techniques for algebra in the ring $\mathbb{Z}\left[\frac{1}{\sqrt{2}}, i\right]$, a 4-qubit unitary over $\mathbb{Z}\left[\frac{1}{\sqrt{2}}, i\right]$ requires specification of the entire unitary, a 16 by 16 matrix where each element can be represented by 5 integers. Without any kind of compression and using 32 bit integers, a single unitary would require $40,960$ bits of memory, a blowup by a factor of 640 compared to reversible functions. To compose circuits, expensive matrix multiplication needs to be performed, and even permutations or inversions need to examine each element of the matrix, adding significant complexity over the bitwise operations for reversible functions. While there may be a more efficient representation of unitary operators that also permits bitwise procedures for permutation, inversion and circuit composition, the authors are not aware of any such representation, and from a strictly information theoretic standpoint it appears to be a much harder problem even with a compact representation.

As an additional point regarding our computational efficiency, we stress that space-time trade-offs were made to allow searches of reasonable depth to be performed. By storing only the circuits, not entire unitaries, we achieved a significant reduction in memory usage, reducing the minimum space to store an $n$-qubit unitary from $4^n\times 5\times 4$ bytes to as few as $n$ bytes. While permutation and inversion is cheap for these circuit descriptions, a circuit of depth $d$ needs $d$ matrix multiplications to compute the corresponding unitary, a major obstacle to performance. To alleviate this time penalty, circuit keys were introduced, storing only a small matrix of complex numbers so that unitaries are only generated when generating a new key, or if a key collision is found.

\subsection{Future work}

The next step in this work will be focused on extending the algorithm to deal with the case when the unitary cannot be implemented exactly. Many common quantum operations are known to be impossible to implement exactly within the Clifford and $T$ gate set, and in particular phase rotations of $e^{2\pi i / 2^m}$, commonly seen in the quantum Fourier transform, are not implementable using Clifford and $T$ gates for $m\geq 4$ \cite{L1}. As a result, a quantum compiler will need efficient approximation sequences for many of the logical operations used in algorithms, and so an algorithm returning depth-optimal $\epsilon$-approximations with the complexity bounds of our exact-searching algorithm would be invaluable. 

Rather than searching for exact collisions within sets $S_i^\dagger U$ and $S_j$, $\epsilon$-approximations of $U$ can be found by examining nearest neighbour pairs between the two sets. While it is unlikely that search times would be as fast as in the exact case, many classical data structures for nearest neighbour searching in metric spaces exist, with some providing complexity bounds comparable to balanced binary trees.

By extending our work to unitary approximations, we could also leverage our advantage over brute force searching to provide a significant speed-up for the Solovay-Kitaev algorithm \cite{D1}. Specifically, Dawson and Nielsen's algorithm \cite{D1} requires the computation of an $\epsilon$-net, in which depth-optimal basic approximations are then looked up when needed -- this phase requires time exponential in the number of qubits. Using our techniques to look up basic approximations could potentially be a major improvement to the algorithm, by allowing these basic approximations to be more accurate and computed in less time.

Another direction for future work will focus on optimizing large circuits using both this algorithm, and databases of circuits generated from this algorithm. While this algorithm performs well for small unitaries, it remains exponential in the size of the instruction set and, by extension, depth, so it is unlikely to be useful for large circuits. Instead, databases generated using these techniques will be instrumental in implementing effective peep-hole optimization and circuit re-synthesis algorithms, or other more scalable approaches to large-scale circuit optimization.

\section{Acknowledgments}
We would like to thank John Watrous and Vadym Kliuchnikov for many useful discussions and contributions.

Supported in part by the Intelligence Advanced Research Projects Activity (IARPA) via Department of Interior National Business Center Contract number DllPC20l66. The U.S. Government is authorized to reproduce and distribute reprints for Governmental purposes notwithstanding any copyright annotation thereon. Disclaimer: The views and conclusions contained herein are those of the authors and should not be interpreted as necessarily representing the official policies or endorsements, either expressed or implied, of IARPA, DoI/NBC or the U.S. Government. 

This material is based upon work partially supported by the National Science Foundation (NSF), during D. Maslov's assignment at the Foundation.
Any opinion, findings, and conclusions or recommendations expressed in this material are those of the author(s) and do not necessarily reflect the views of the National Science Foundation.

Michele Mosca is also supported by Canada’s NSERC, MITACS, CIFAR, and CFI.

IQC and Perimeter Institute are supported in part by the Government of Canada and the Province of Ontario.

All circuit figures in this paper were generated using QCViewer \cite{QCV}.


\begin{thebibliography}{2}

\bibliographystyle{IEEEtran}

\bibitem[1]{G2} S. Aaronson, D. Gottesman, \emph{Improved simulation of stabilizer circuits}. Phys. Rev. A 70, 052328, 2004, \href{http://arxiv.org/abs/quant-ph/0406196}{quant-ph/0406196}.

\bibitem[2]{G1} P. Aliferis, D. Gottesman, J. Preskill, \emph{Quantum accuracy threshold for concatenated distance-3 codes}. Quantum Information and Computation {\bf 6}:97--165, 2006, \href{http://arxiv.org/abs/quant-ph/0504218}{quant-ph/0504218}.

\bibitem[3]{BS} A. Bocharov, K. M. Svore, \emph{A Depth-Optimal Canonical Form for Single-qubit Quantum Circuits}. \href{http://arxiv.org/abs/1206.3223}{arXiv:1206.3223}, 2012.

\bibitem[4]{B1} H. Bombin {\it et al.}, \emph{Strong resilience of topological codes to depolarization}. Phys. Rev. X 2, 021004, 2012, \href{http://arxiv.org/abs/1202.1852}{arXiv:1202.1852}.

\bibitem[5]{QS} J. W. Britton {\it et al.}, \emph{Engineered two-dimensional Ising interactions in a trapped-ion quantum simulator with hundreds of spins}. Nature {\bf 484}:489-–492, 2012.

\bibitem[6]{SQ} K. R. Brown {\it et al.}, \emph{Single-qubiT gate error below $10^{-4}$ in a trapped ion}. Phys. Rev. A 84, 030303(R), 2011, \href{http://arxiv.org/abs/1104.2552/}{arXiv:1104.2552}.

\bibitem[7]{WELD} A. Childs {\it et al.}, \emph{Exponential algorithmic speedup by quantum walk}. Proc. 35th ACM Symposium on Theory of Computing, pages 59--68, 2003, \href{http://arxiv.org/abs/quant-ph/0209131}{quant-ph/0209131}.

\bibitem[8]{IBM1} J. M. Chow {\it et al.}, \emph{Complete universal quantum gate set approaching fault-tolerant thresholds with superconducting qubits}. \href{http://arxiv.org/abs/1202.5344}{arXiv:1202.5344}, 2012.

\bibitem[9]{D1} C. Dawson, M. Nielsen, \emph{The Solovay-Kitaev algorithm}. Quantum Information and Computation {\bf 6}:81--95, 2006, \href{http://arxiv.org/abs/quant-ph/0505030}{quant-ph/0505030}.

\bibitem[10]{FEYN} R. P. Feynman, \emph{Quantum mechanical computers}. Foundations of Physics {\bf 16}(6):507--531, 1986.

\bibitem[11]{F1} A. Fowler, \emph{Constructing arbitrary Steane code single logical qubit fault-tolerant gates}. Quantum Information and Computation {\bf 11}:867--873, 2011, \href{http://arxiv.org/abs/quant-ph/0411206}{quant-ph/0411206}.

\bibitem[12]{F2} A. G. Fowler, A. M. Stephens, P. Groszkowski, \emph{High threshold universal quantum computation on the surface code}. Phys. Rev. A 80, 052312, 2009, \href{http://arxiv.org/abs/0803.0272}{arXiv:0803.0272}.

\bibitem[13]{F3} A. G. Fowler, A. C. Whiteside, L. C. L. Hollenberg, \emph{Towards practical classical processing for the surface code}. Phys. Rev. Lett. 108, 180501, 2012, \href{http://arxiv.org/abs/1110.5133}{arXiv:1110.5133}.

\bibitem[14]{M1} O. Golubitsky, D. Maslov, \emph{A study of optimal 4-bit reversible Toffoli circuits and their synthesis}. IEEE Transactions on Computers {\bf 61}(9):1341--1353, 2012, \href{http://arxiv.org/abs/1103.2686}{arXiv:1103.2686}.

\bibitem[15]{HUNG} W. N. N. Hung {\it et al.}, \emph{Optimal synthesis of multiple output Boolean functions using a set of quantum gates by symbolic reachability analysis}. IEEE Transactions on CAD {\bf 25}(9):1652--1663, 2006.

\bibitem[16]{KLM} P. Kaye, R. Laflamme, M. Mosca, \emph{An Introduction to Quantum Computing}. Oxford University Press, 2007.

\bibitem[17]{V1} V. Kliuchnikov, D. Maslov, M. Mosca, \emph{Fast and efficient exact synthesis of single qubit unitaries generated by Clifford and T gates}. \href{http://arxiv.org/abs/1206.5236}{arXiv:1206.5236}, 2012.

\bibitem[18]{M2} D. Maslov, D. M. Miller, \emph{Comparison of the cost metrics for reversible and quantum logic synthesis}. IET Computers \& Digital Techniques, {\bf 1}(2):98--104, 2007, \href{http://arxiv.org/abs/quant-ph/0511008}{quant-ph/0511008}.

\bibitem[19]{SK} M. Nielsen, I. L. Chuang, \emph{Quantum Computation and Quantum Information}. Cambridge University Press, 2000.

\bibitem[20]{P1} K. N. Patel, I. L. Markov, J. P. Hayes, \emph{Efficient Synthesis of Linear Reversible Circuits}. \href{http://arxiv.org/abs/quant-ph/0302002}{quant-ph/0302002}, 2003.

\bibitem[21]{PAUL} P. Pham, \emph{Quantum Compiler}. Available at \href{http://sourceforge.net/p/quantumcompiler/home/Home/}{\texttt{http://sourceforge.net/p/quantumcompiler/home/Home/}}, ver. 0.02, 2011.

\bibitem[22]{PER} A. Peres, \emph{Reversible logic and quantum computers}. Phys. Rev. A 32:3266--3276, 1985.

\bibitem[23]{QCV} QCViewer: a tool for displaying, editing, and simulating 
quantum circuits. Available at \href{http://qcirc.iqc.uwaterloo.ca/index.php?n=Projects.QCViewer}{\texttt{http://qcirc.iqc.uwaterloo.ca/}}, ver. 0.8, May 2012.

\bibitem[24]{IBM2} C. Rigetti {\it et al.}, \emph{Superconducting qubit in waveguide cavity with coherence time approaching 0.1ms}, \href{http://arxiv.org/abs/1202.5533}{arXiv:1202.5533}, 2012.

\bibitem[25]{SEL} P. Selinger, private communication, July 16, 2012.

\bibitem[26]{T1} S. Sen, R. E. Tarjan, \emph{Deletion without rebalancing in balanced binary trees}. Proc. 21st ACM-SIAM Symposium on Discrete Algorithms (SODA), pages 1490--1499, 2010.

\bibitem[27]{MARKOV} V. V. Shende {\it et al.}, \emph{Synthesis of reversible logic circuits}.  IEEE Transactions on CAD, {\bf 22}(6):710--722, 2003, \href{http://arxiv.org/abs/quant-ph/0207001}{quant-ph/0207001}.

\bibitem[28]{L1} X. Zhou, D. W. Leung, I. L. Chuang, \emph{Methodology for quantum logic gate constructions}. Phys. Rev. A 62, 052316, 2000, \href{http://arxiv.org/abs/quant-ph/0002039}{quant-ph/0002039}.

\end{thebibliography}
\end{document}